\documentclass[journal]{IEEEtran}
\usepackage{array}	
\usepackage[nolist]{acronym}
\usepackage[utf8]{inputenc}
\usepackage[T1]{fontenc}
\usepackage{multirow}
\usepackage{rotating}
\usepackage{url}
\usepackage{graphicx}
\usepackage[dvipsnames]{xcolor}
\ifCLASSOPTIONcompsoc
  \usepackage[caption=false,font=normalsize,labelfont=sf,textfont=sf]{subfig}
\else
  \usepackage[caption=false,font=footnotesize]{subfig}
\fi
\graphicspath{{../figs/}}
\usepackage{amsmath, mathtools, stmaryrd, amssymb}
\usepackage[noadjust]{cite}	

\newcommand{\Ry}{\mathbf{R}_{yy}}

\newcommand{\w}{\mathbf{w}}

\newcommand{\y}{\mathbf{y}}

\newcommand{\s}{\mathbf{s}}
\newcommand{\n}{\mathbf{n}}
\newcommand{\g}{\mathbf{g}}
\newcommand{\z}{\mathbf{z}}

\begin{document}
\begin{acronym}
\acro{ds}[DSB]{delay-and-sum beamformer}
\acro{mpdr}[MPDR]{minimum power distortionless response beamformer}
\acro{mvdr}[MVDR]{minimum variance distortionless response beamformer}
\acro{lcmp}[LCMP]{linearly constrained minimum power beamformer}
\acro{lcmv}[LCMV]{linearly constrained minimum variance beamformer}
\acro{mwf}[MWF]{multichannel Wiener filter}
\acro{sdw}[SDW-MWF]{speech distortion weighted multichannel Wiener filter}
\acro{mvdr}[MVDR]{minimum variance distortionless response}
\acro{gevd}[GEVD]{generalized eigenvalue decomposition}
\acro{nmf}[NMF-MWF]{non-negative matrix factorization}
\acro{stft}[STFT]{short-time Fourier transform}
\acro{tf}[TF]{time-frequency}
\acro{vad}[VAD]{voice activity detector}
\acroplural{vad}[VADs]{voice activity detectors}
\acro{danse}[DANSE]{distributed adaptive node-specific signal estimation}
\acro{mse}[MSE]{mean squared error}
\acro{wasn}[WASN]{wireless acoustic sensor network}
\acroplural{wasn}[WASNs]{wireless acoustic sensor networks}
\acro{doa}[DOA]{direction of arrival}
\acroplural{doa}[DOAs]{directions of arrival}
\acro{irm}[IRM]{ideal ratio mask}
\acroplural{irm}[IRMs]{ideal ratio masks}
\acro{ibm}[IBM]{ideal binary mask}
\acro{dnn}[DNN]{deep neural network}
\acroplural{dnn}[DNNs]{deep neural networks}
\acro{nn}[NN]{neural network}
\acroplural{nn}[NNs]{neural networks}
\acro{lstm}[LSTM]{long short-term memory}
\acro{cnn}[CDNN]{convolutional neural network}
\acroplural{cnn}[CNNs]{convolutional neural networks}
\acro{gru}[GRU]{gated recurrent unit}
\acro{crnn}[CRNN]{convolutional recurrent neural network}
\acro{rnn}[RNN]{recurrent neural network}
\acroplural{rnn}[RNNs]{recurrent neural networks}
\acro{rir}[RIR]{room impulse response}
\acroplural{rir}[RIRs]{room impulse responses}
\acro{ssn}[SSN]{speech shaped noise}
\acro{snr}[SNR]{signal to noise ratio}
\acroplural{snr}[SNRs]{signal to noise ratios}
\acro{sar}[SAR]{source to artifacts ratio}
\acro{sir}[SIR]{source to interferences ratio}
\acroplural{sir}[SIRs]{source to interferences ratios}
\acro{sdr}[SDR]{source to distortion ratio}

\end{acronym}

%
\title{DNN-based mask estimation for distributed speech enhancement in spatially unconstrained microphone arrays}
%
%
%

\author{Nicolas~Furnon,~Romain~Serizel,~Irina~Illina,~Slim~Essid%
\thanks{N. Furnon, R. Serizel and I. Illina are with the Université de Lorraine, CNRS, Inria, Loria, F-54000 Nancy, France. email: nicolas.furnon@loria.fr}%
\thanks{Slim Essid is with the LTCI, Télécom Paris, Institut Polytechnique de Paris, Palaiseau, France.}}

\markboth{Journal of \LaTeX\ Class Files,~Vol.~14, No.~8, August~2015}%
{Shell \MakeLowercase{\textit{et al.}}: DNN-based mask estimation for distributed speech enhancement in spatially unconstrained microphone arrays}
%



\maketitle

\begin{abstract}
Deep neural network (DNN)-based speech enhancement algorithms in microphone arrays have now proven to be efficient solutions to speech understanding and speech recognition in noisy environments. However, in the context of ad-hoc microphone arrays, many challenges remain and raise the need for distributed processing. In this paper, we propose to extend a previously introduced distributed DNN-based time-frequency mask estimation scheme that can efficiently use spatial information in form of so-called compressed signals which are pre-filtered target estimations. We study the performance of this algorithm under realistic acoustic conditions and investigate practical aspects of its optimal application. We show that the nodes in the microphone array cooperate by taking profit of their spatial coverage in the room. We also propose to use the compressed signals not only to convey the target estimation but also the noise estimation in order to exploit the acoustic diversity recorded throughout the microphone array.

\end{abstract}

%
\IEEEpeerreviewmaketitle

\section{Introduction}
\IEEEPARstart{S}{peech} enhancement aims to recover the clean speech from a noisy signal. It can be used in applications as diverse as automatic speech recognition, hearing aids, and (hand-free) mobile communication. Single-channel speech enhancement, relying on a single microphone signal, can substantially increase the speech quality but the noise reduction is often accompanied by an increase in the speech distortion. 
Multichannel speech enhancement can overcome this limitation by exploiting the spatial information provided by several microphones. One can distinguish the data-independent multichannel filters \cite{VanVeen1988} from the data-dependent multichannel filters  \cite{Capon1969, Frost1972, VanTrees2002, Doclo2002}, which depend on the estimation of the statistics of the noisy signal, the noise signal or the target signal. The \ac{mwf} is a data-dependent multichannel filter, which is optimal in the \ac{mse} sense. It can be extended to the \ac{sdw} \cite{Doclo2007} which enables a trade-off between the noise reduction and the speech distortion.
Most of these multichannel filters have been developed in constrained microphone arrays, where the number and positions of microphones are fixed and where all the microphones share a common clock. They are called centralized solutions, because a so-called fusion center gathers all the signals of the microphone array.

With the multiplication of embedded microphones in wireless portable devices that surround us, ad-hoc microphone arrays have gained interest \cite{Gannot2017}. They can be considered as heterogeneous, unconstrained microphone arrays, which are much more flexible and can cover a wider area than traditional microphone arrays. However, the dependency of the centralized approaches on a fusion center makes these solutions too constrained and unrealistic. Many solutions have been proposed to distribute the processing over the whole microphone array in order to get rid of the fusion center, based on a reduction of the transmission costs \cite{Roy2009, JZhang2018a, Bertrand2010} or on distributed processing \cite{Zeng2013, Zheng2012, GZhang2018, Koutrouvelis2018}. Bertrand and Moonen introduced a distributed version of the \ac{mwf}, where each node, instead of sending all its signals to a fusion center, sends only one signal, called compressed signal, to the other nodes, thus reducing the bandwidth cost in addition to cancelling the need of a fusion center \cite{Bertrand2010a}.

All of these methods rely on the knowledge either of the (relative) acoustic transfer functions, or of the target signals covariance matrices, or both. Recently, \ac{dnn}-based solutions have enabled great progress to accurately estimate these parameters, most of the time by predicting \ac{tf} masks from a single-channel input \cite{Heymann2016, Wang2018, Erdogan2016}. However, it is also possible to exploit the multichannel information to better estimate these parameters. The spatial information can be explicitly given to a \ac{dnn} through handcrafted features \cite{Jiang2014}, or implicitly by feeding the \ac{dnn} either with the multichannel \ac{stft} signals \cite{Perotin2018a, Yoshioka2018, Chakrabarty2019} or with the multichannel raw waveforms \cite{Sainath2017}.

Although these \ac{dnn}-based methods lead to promising results and manage to exploit multichannel information, most of them are centralized solutions. Very little work has been published on \ac{dnn}-based speech enhancement in ad-hoc microphone arrays. Ceolini and Liu \cite{Ceolini2019} introduced a \ac{dnn}-based method that can process real-time speech enhancement in an ad-hoc microphone array but their solution relies on a centralized \ac{mvdr} and the \ac{dnn} is not able to exploit multichannel information. 
In a previously published paper, we introduced a distributed \ac{dnn}-based mask estimation that could exploit the multichannel data to better predict the masks \cite{Furnon2020}. 
Tested in a simple scenario with two nodes, it outperformed a \ac{mwf} applied to the nodes separately. 

This paper proposes an extended study of our previously introduced speech enhancement scheme \cite{Furnon2020}. By analysing its performance under various configurations, including real world ones, we confirm that it matches \ac{danse} performance with an oracle \ac{vad}. We also evaluate in detail the performance at the different nodes in the microphone array in order to highlight their cooperation. This study shows that, depending on the characteristics of the signals captured by the sending and the receiving nodes, sending the so-called compressed signal could be optimized by deciding to send the estimation of either the target or the noise.

Besides, we analyse the performance of the \ac{dnn} used in this context. In particular, we investigate the influence of the noise and the spatial diversity between the training and test conditions, looking for a trade-off between performance and robustness to varying scenarios. We also investigate the influence of the quality (in terms of \ac{sir}) of the signals used to train the \acp{dnn}.

This paper is organized as follows. In Section \ref{sec:formulation}, we describe the problem and the multichannel speech enhancement solutions that this paper relies on. In Section \ref{sec:contribution} we present our proposal and the challenges that it raises. The experimental setup used to evaluate our proposed solution is described in Section \ref{sec:setup}. In Sections \ref{sec:optim_sn} and \ref{sec:optim_mn}, we investigate the performance of the \acp{dnn} which have single-channel and multi-channel input. We show in Section \ref{sec:zs_zn} that sending the noise estimation can lead to improved performance. We conclude the paper in Section \ref{sec:conclusion}.

\section{Problem formulation}
\label{sec:formulation}

\subsection{Notations}
We consider a fully-connected microphone array with $K$ nodes each having $M_k$ microphones. $M=\sum_{k=1}^{K} M_k$ is the total number of microphones. The signal recorded by the $m$-th microphone of the $k$-th node is denoted as $y_{k, m}$. Under the assumption of an additive noise model, in the \ac{stft} domain, we have:
\[
y_{k, m}(f, t)~=~s_{k, m}(f, t) + n_{k, m}(f, t)\,,
\] 
where $s_{k, m}$ and $n_{k, m}$ denote the speech and noise signals respectively and where $f$ and $t$ denote the frequency and time frame indexes respectively. For the sake of conciseness, we will thereafter omit the frame and frequency indexes unless necessary. The signals from the different channels at node $k$ are stacked into the vector:
\[
\y_k~=~[y_{k, 1}, ..., y_{k, M_k}]^T.
\]
All the signals of all nodes are stacked into the vector $\y~=~[\y_{1}^T, ..., \y_{K}^T]^T$. Similarly, the speech and noise signals are stacked into $\s$ and $\n$. In the following, regular lowercase letters denote scalars; bold lowercase letters indicate vectors and bold uppercase letters indicate matrices.

\subsection{Multichannel Wiener filter}
The centralized \ac{mwf} aims at estimating the speech component $s_{i}$ of the $i$-th sensor of the microphone array. The \ac{mwf} is the optimal filter in the \ac{mse} sense, \textit{i.e.} it minimises the \ac{mse} between the desired signal $s_i$ and the estimated signal:
\begin{equation}
\label{eq:mwf}
	\w_{\mathrm{MWF}} = \mathrm{arg}\min_{\w} \mathbb{E}\{|s_{i} - \w^H\y|^2\}.
\end{equation}

$\mathbb{E}\{\cdot\}$ is the expectation operator and $\cdot^H$ denotes the Hermitian transpose.
Solving Eq. \eqref{eq:mwf} yields:

\begin{equation}
	\w_{\mathrm{MWF}} = \Ry^{-1}\mathbf{R}_{ys}\mathbf{e}_i\,,
\end{equation}
where $\Ry$ is the correlation matrix of the input signal, $\mathbf{R}_{ys}$ is the cross-correlation matrix between the input signal and its speech component and $\mathbf{e}_i \in \mathbb{R}^{M}$ is a vector of zeros with a $1$ at the $i$-th position. Without loss of generality, we will take the channel $i=1$ as the reference channel in the sequel. The correlation matrices can be obtained as follows:
\begin{eqnarray}
\Ry &=& \mathbb{E}\{\y\y^H\} \\
\mathbf{R}_{ys} &=& \mathbb{E}\{\y\s^H\}\,.
\end{eqnarray}

Under the assumption that speech and noise are uncorrelated and that the noise is locally stationary, we have:
\begin{equation}\label{eq:rss_basic}
	\mathbf{R}_{ys} = \mathbf{R}_{ss} = \mathbb{E}\{\mathbf{s}\mathbf{s}^H\} = \mathbf{R}_{yy} - \mathbf{R}_{nn}
\end{equation} 
where $\mathbf{R}_{nn}$ is the noise correlation matrix:
\begin{equation}
\mathbf{R}_{nn} = \mathbb{E}\{\mathbf{n}\mathbf{n}^H\}\,.
\end{equation}
Computing these matrices requires the knowledge of noise-only periods and speech-plus-noise periods. This is typically obtained with a \ac{vad} \cite{Doclo2007, Bertrand2010a} or a \ac{tf} mask \cite{Heymann2016, Ceolini2019, Pfeifenberger2017}.

Under the assumption that a single speech source is present, Serizel et al. proposed a rank-1 approximation of the covariance matrix $\mathbf{R}_{ss}$ based on the \ac{gevd} of the matrix pencil $\{\mathbf{R}_{yy},\mathbf{R}_{nn}\}$ \cite{Serizel2014}: 
\begin{align}
	\mathbf{R}_{nn} &=\mathbf{Q} \boldsymbol{\Sigma}_{n} \mathbf{Q}^{H} \label{subeq:rnn_gevd}\\
	\mathbf{R}_{yy} &=\mathbf{Q} \boldsymbol{\Sigma}_{y} \mathbf{Q}^{H} \label{subeq:ryy_gevd}
\end{align}
where $\mathbf{Q}$ is the matrix of the generalized eigenvectors. $\boldsymbol{\Sigma}_y~=~\text{diag}\{\sigma_{y_1} \dots\sigma_{y_M}\}$ and $\boldsymbol{\Sigma}_n~=~\text{diag}\{\sigma_{n_1}\dots\sigma_{n_M}\}$ are the diagonal matrices of the generalized eigenvalues. Plugging \eqref{subeq:rnn_gevd} and \eqref{subeq:ryy_gevd} into \eqref{eq:rss_basic} gives:
\[\mathbf{R}_{ss}=\mathbf{Q}(\underbrace{\mathbf{\Sigma}_{y}-\mathbf{\Sigma}_{n}}_{\Sigma_{s}}) \mathbf{Q}^{H}\]
where $\boldsymbol{\Sigma}_s~=~\text{diag}\{\sigma_{s_1}\dots\sigma_{s_M}\}$. $\mathbf{R}_{ss}$ can be approximated by a rank-1 decomposition:
\begin{equation}
	\label{eq:r1_q1}
	\mathbf{R}_{ss}\approx\underbrace{\mathbf{q}_{1} \mathbf{q}_{1}^{H} \sigma_{s_{1}}}_{\mathbf{R}_{s_{r 1}}}
\end{equation}
where $\mathbf{q}_1$ is the first column of $\mathbf{Q}$. We now consider $\mathbf{t}_{1}~=~\mathbf{Q}^{-H} \mathbf{e}_{1} \mathbf{q}_{1}(1)^{*}$, the projector into the space spanned by $\mathbf{q}_1$ with $\mathbf{q}_1(1)$ the first element of $\mathbf{q}_1$ and $\cdot^*$ the complex conjugate operator.
Replacing the desired signal $s_1 = \mathbf{e}_1^T\y$ in \eqref{eq:mwf} by the implicit reference $\mathbf{t}_1^H\y$, and using the \ac{sdw} extension of the \ac{mwf} \cite{Doclo2007} leads to the new cost function:
\begin{align}\label{eq:j_gevd_sdw}
	J_{\mathrm{GEVD}-\mathrm{SDW}-\mathrm{MWF}} &=\mathbb{E}\left\{\left|\mathbf{w}^{H} \mathbf{s} - \mathbf{t}_{1}^H\mathbf{y} \right|^{2}\right\} \nonumber\\
	&+\mu \mathbb{E}\left\{\left|\mathbf{w}^{H} \mathbf{n}\right|^{2}\right\}
\end{align}
with $\mu$ a trade-off parameter between the noise reduction and the speech distortion. The solution to \eqref{eq:j_gevd_sdw} is given by:
\begin{equation}\label{eq:w_gevd_sdw}
	\mathbf{w}_{\mathrm{GEVD}-\mathrm{SDW}-\mathrm{MWF}}=\left(\mathbf{R}_{s_{r 1}}+\mu \mathbf{R}_{nn}\right)^{-1} \mathbf{R}_{s_{r 1}} \mathbf{e}_{1}\,.
\end{equation}
The resulting filter proved to be more robust in low \ac{snr} scenarios with a stronger noise reduction \cite{Serizel2014}.

\subsection{Distributed multichannel Wiener filter}
\label{subsec:danse}
The \ac{danse} algorithm is a distributed \ac{mwf} which aims at estimating the speech component $s_{k, i}$ of the $i$-th microphone of every node $k$ \cite{Bertrand2010a, Hassani2015}. We still assume that a single speech source is present. In the \ac{danse} algorithm, no fusion center gathers all the signals of all nodes. Instead, every node $k$ sends only one so-called compressed signal $z_k$ to the other nodes and receives $K-1$ signals from the other nodes. A \ac{sdw} is applied to the vector
\begin{eqnarray} \tilde{\y}_k~=~\left[ \y_k^T,~\z_{-k}^T\right]^T \end{eqnarray}
\begin{center} where $\z_{-k}~=~[z_1, ..., z_{k-1}, z_{k+1}, ..., z_K]^T$ \end{center}
and outputs the estimated speech signal $\hat{s}_k$ as follows:

\begin{align}
	\label{eq:danse_a}
	\hat{s}_k 	&= \w_k^H \tilde{\y}_k \\
	\label{eq:danse_b}
	~		&= \w_{kk}^H\y_k + \g_{-k}^H\z_{-k}\,,
\end{align}
where $\w_k = \left[ \w_{kk}^T,~\g_{-k}^T\right]^T$ is the so-called global filter. $\w_{kk}$ and $\g_{-k}$ are filters applied on the noisy signal $\y_k$ and the stacked compressed signals $\z_{-k}$ respectively.
Similarly to Eq.~\eqref{eq:w_gevd_sdw}, it can be computed as
\begin{equation}
	\label{eq:danse_filt}
	\w_{k} = \big(\mathbf{R}_{s_{r 1}, k} + \mu\mathbf{R}_{nn, k}\big)^{-1}\mathbf{R}_{s_{r 1}, k}\mathbf{e}_1\,,
\end{equation} 
where $\mathbf{R}_{s_{r 1}, k}$ and $\mathbf{R}_{nn, k}$ are estimated from $\tilde{\y}_k$. 
From Eq.~\eqref{eq:danse_b}, it can be seen that the sub-filter $\w_{kk}$ is applied on the local signals $\y_k$ only, which yields the compressed signal $z_k$ to be sent to the other nodes:
\begin{equation}\label{eq:z_k}
	z_k = \w_{kk}^H\y_k\,.
\end{equation}

\subsection{Mask-based multichannel speech enhancement}
\label{subsec:mask_based_se}
Originally, the \ac{dnn}-predicted \ac{tf} masks were directly applied to the \ac{stft} of the noisy signal in order to extract the target speech \cite{Weintraub1985, Jin2009}. This idea continues to be used with a good performance both in the single-channel \cite{Li2020} and the multichannel context \cite{Gu2020}, but it requires much better \ac{tf} masks and complex \ac{dnn} architectures. It also suffers from distortion that can be alleviated by using multichannel filters. In microphone arrays, a common practice is to estimate a \ac{tf} mask that is not directly applied to the noisy signal, but used to replace the \ac{vad} necessary to compute the speech and noise statistics \cite{Frost1972, VanTrees2002, Doclo2002} required by the multichannel filters like \ac{mvdr} \cite{Heymann2016, Ceolini2019} or \ac{mwf} \cite{Pfeifenberger2017}. Using these \ac{tf} masks, the speech covariance matrix can be estimated as:
\begin{equation}
\label{eq:rss}
\mathbf{R}_{ss, k} = \mathbb{E}\{\mathbf{\tilde{s}}_k\mathbf{\tilde{s}}^H_k\} 
\end{equation}
with 
\begin{equation}
\mathbf{\tilde{s}}_k(t, f) = \mathbf{m}_{s,k}(t, f)\odot\mathbf{\tilde{y}}_k(t, f)
\end{equation}
where $\odot$ is the Hadamard product and $\mathbf{m}_{s,k}$ are the stacked \ac{tf} masks corresponding to the speech components of $\mathbf{\tilde{y}}_k$. To compute the noise covariance matrix, the \ac{tf} masks $\mathbf{m}_{s,k}$ should be replaced by their complement $\mathbf{m}_{n,k} = 1 - \mathbf{m}_{s,k}$.

\begin{figure}
	\centering
	\includegraphics[width=1\linewidth]{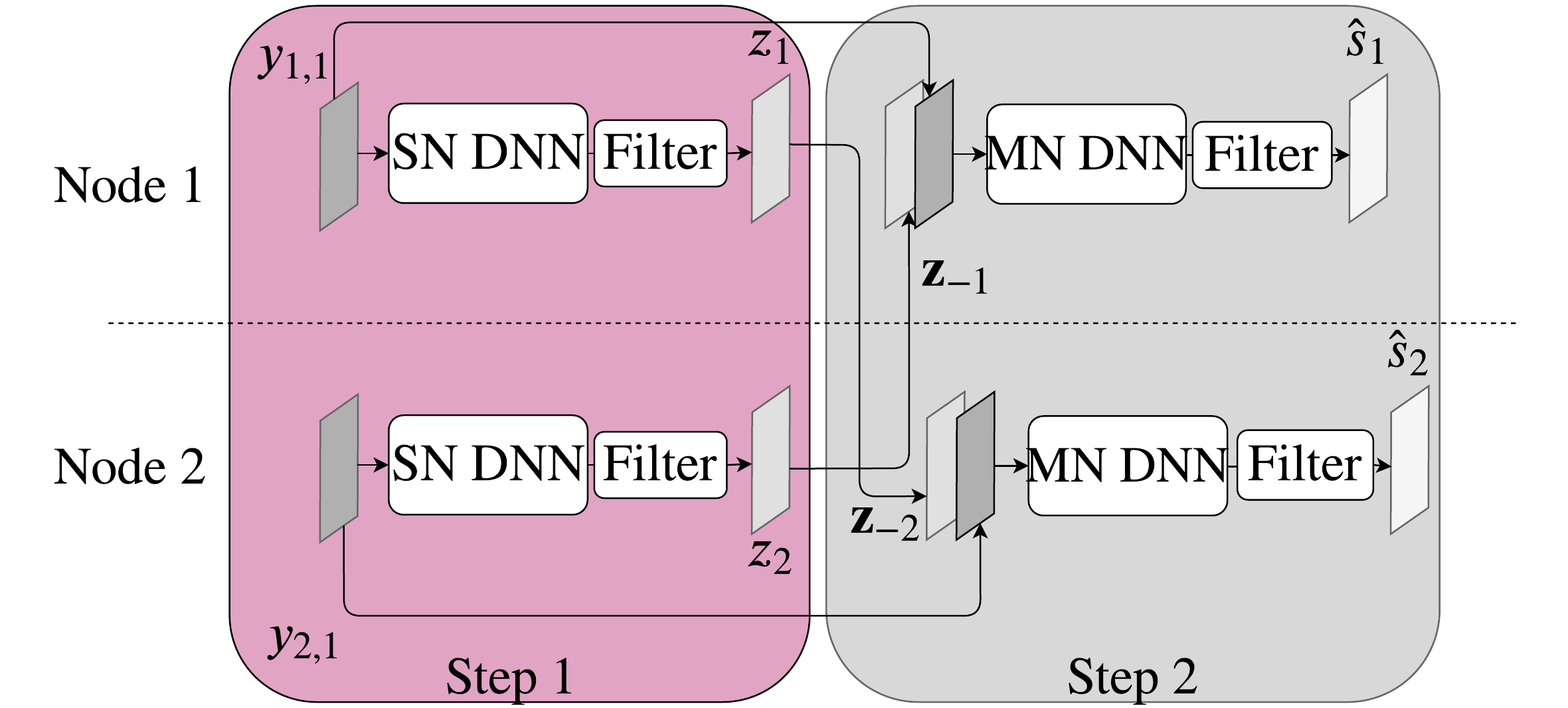}
	\caption{Two-step version of the DANSE algorithm in a context with two nodes. "SN DNN" and "MN DNN" respectively refer to single-node and multi-node deep neural networks.}
	\label{fig:tango_nn}
\end{figure}

\begin{figure}
	\centering
	\includegraphics[width=\linewidth]{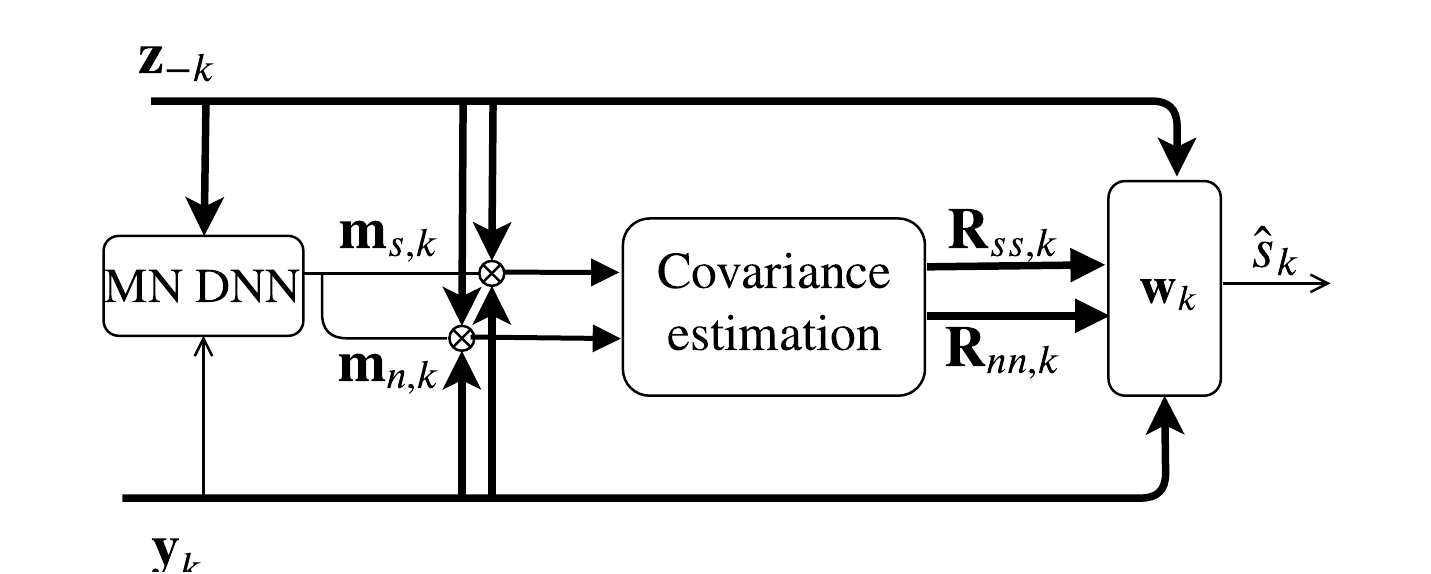}
	\caption{Detail of the second filtering step. Bold arrows represent multichannel signals, simple ones represent single-channel signals.}
	\label{fig:step2}
\end{figure}

To estimate these \ac{tf} masks, a common practice is to use a \ac{dnn} which estimates them from a single-channel noisy signal~\cite{Heymann2016, Pfeifenberger2017, Ceolini2019}. In a previous work, we showed that we can improve the \ac{tf} mask prediction, as represented in Figure~\ref{fig:tango_nn}~\cite{Furnon2020}. In a two-node scenario, we introduced a batch-version of \ac{danse} where at each node, a \ac{crnn} predicted a \ac{tf} mask out of the reference channel of the node and the compressed signal sent by the other node.
To avoid issues related to convergence, we split the iterative process of \ac{danse} into two distinct steps. In a first step (left box of Figure~\ref{fig:tango_nn}), each node processed only local signals to estimate the compressed signal as $z_k = \w_{kk}^H\y_k$ and sent it. In a second step, detailed in Figure~\ref{fig:step2}, each node used both local and compressed signals to estimate the desired signal. As in the original version of \ac{danse}, the compressed signal is used to compute the speech and noise covariance matrices, but we additionally use it to better predict the \ac{tf} mask with the multi-node \ac{dnn}.

In the rest of the paper, we will refer as \textit{single-node} \acp{dnn} to the \acp{dnn} which predict a \ac{tf} mask based on the signal of only one node (\textit{e.g.} the \ac{dnn} of the first step in Figure~\ref{fig:tango_nn}), and as \textit{multi-node} \acp{dnn} to the \acp{dnn} which predict a \ac{tf} mask based on signals coming from several nodes (\textit{e.g.} the \ac{dnn} of the second step in Figure~\ref{fig:tango_nn}).

\section{Analysis the DNN-based distributed multichannel Wiener filter}
\label{sec:contribution}
\begin{figure*}[!t]
	\centering
	\subfloat[]{\includegraphics[width=2.4in]{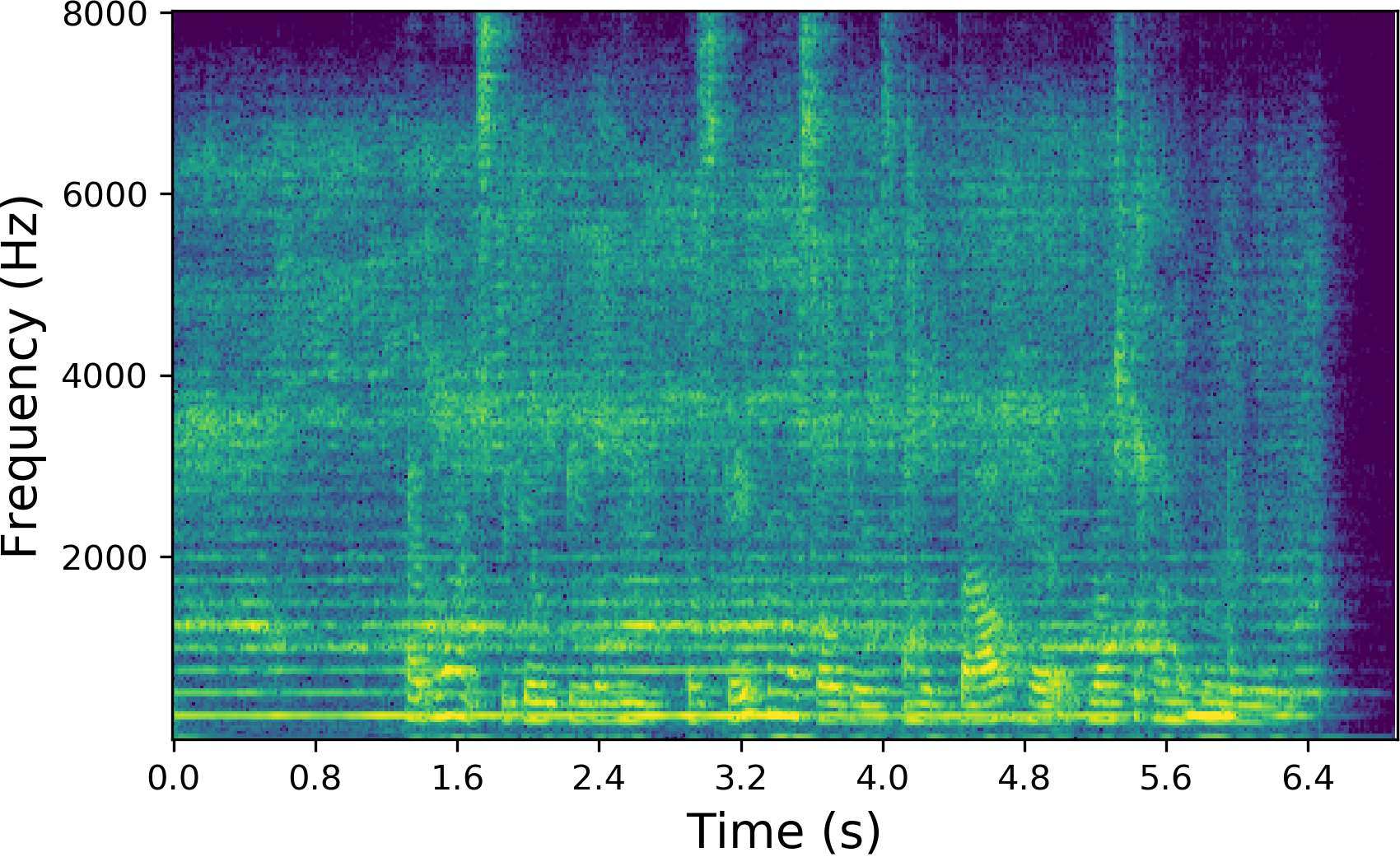}%
		\label{subfig:input}}
	\subfloat[]{\includegraphics[width=2.4in]{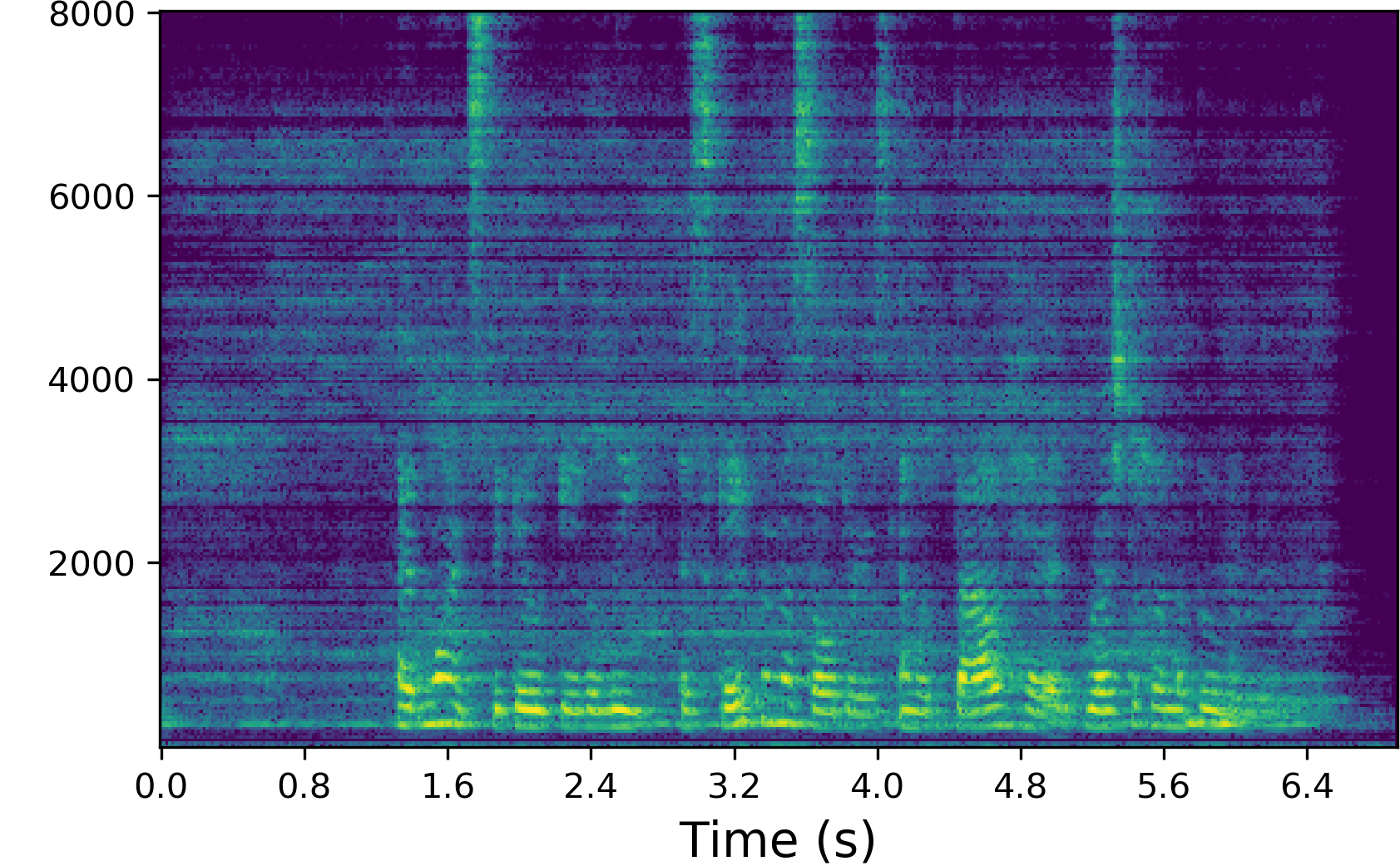}%
		\label{subfig:z}}
	\subfloat[]{\includegraphics[width=2.4in]{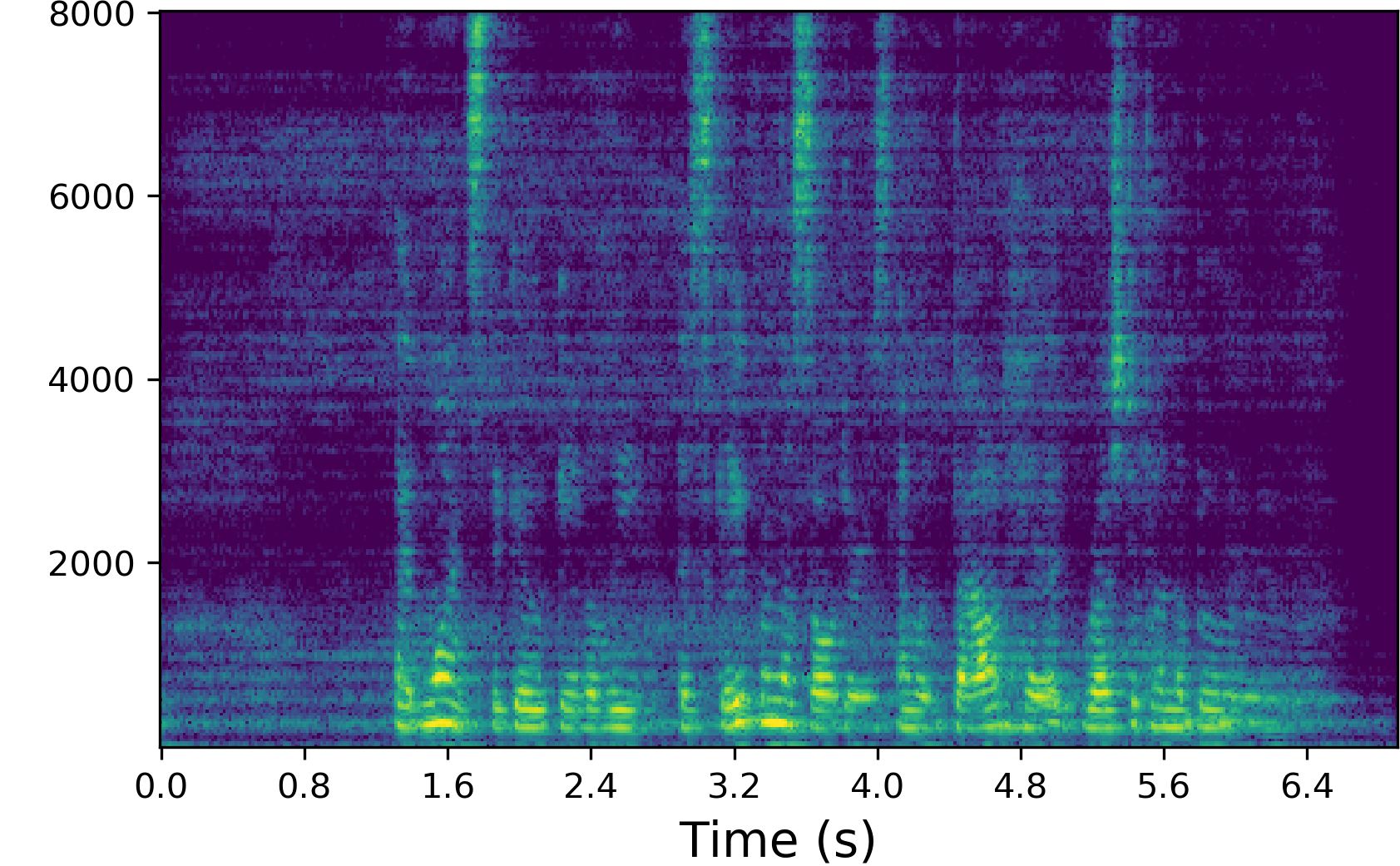}%
		\label{subfig:output}}
	\hfil
	\subfloat[]{\includegraphics[width=2.4in]{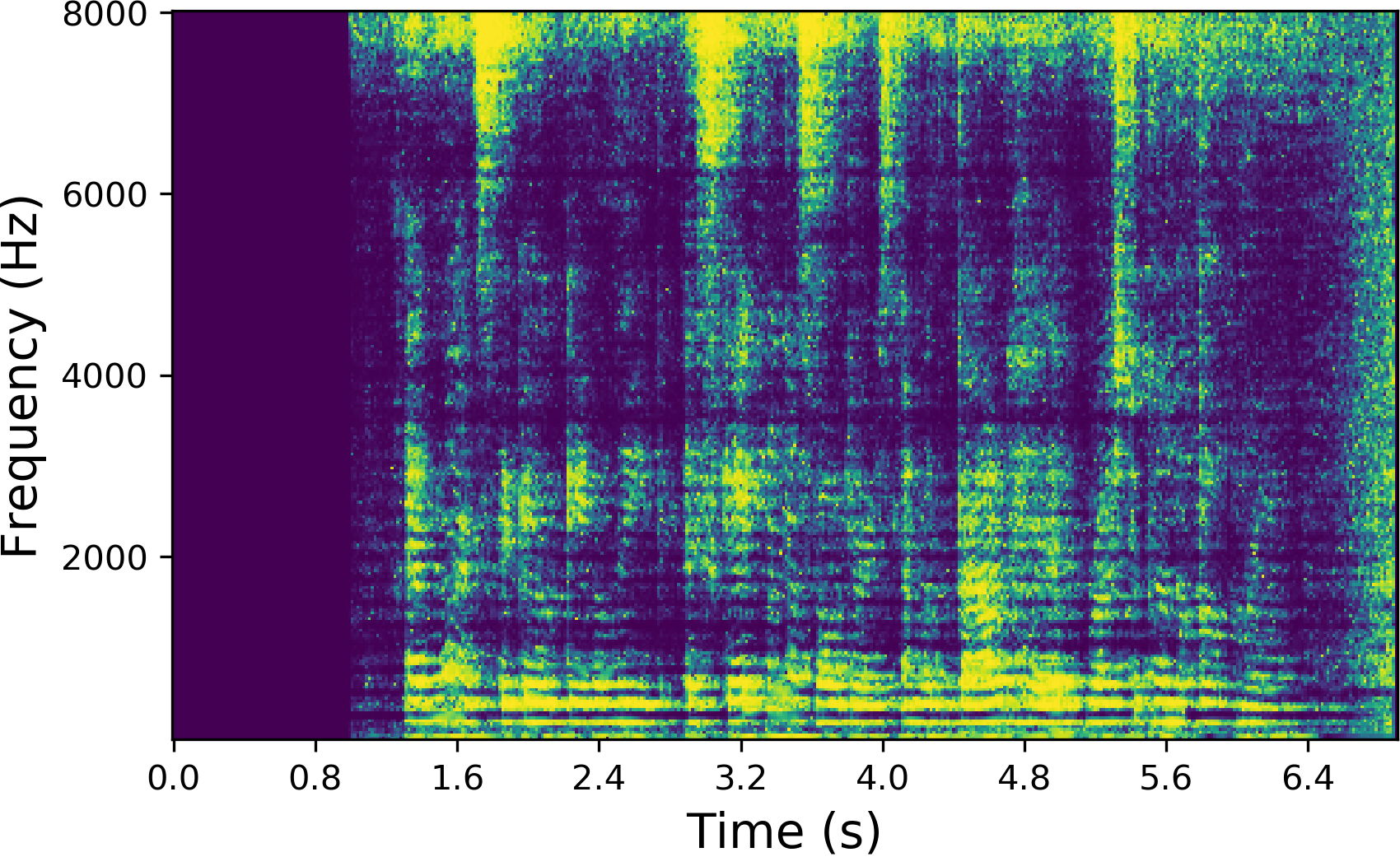}%
		\label{subfig:irm}}
	\subfloat[]{\includegraphics[width=2.4in]{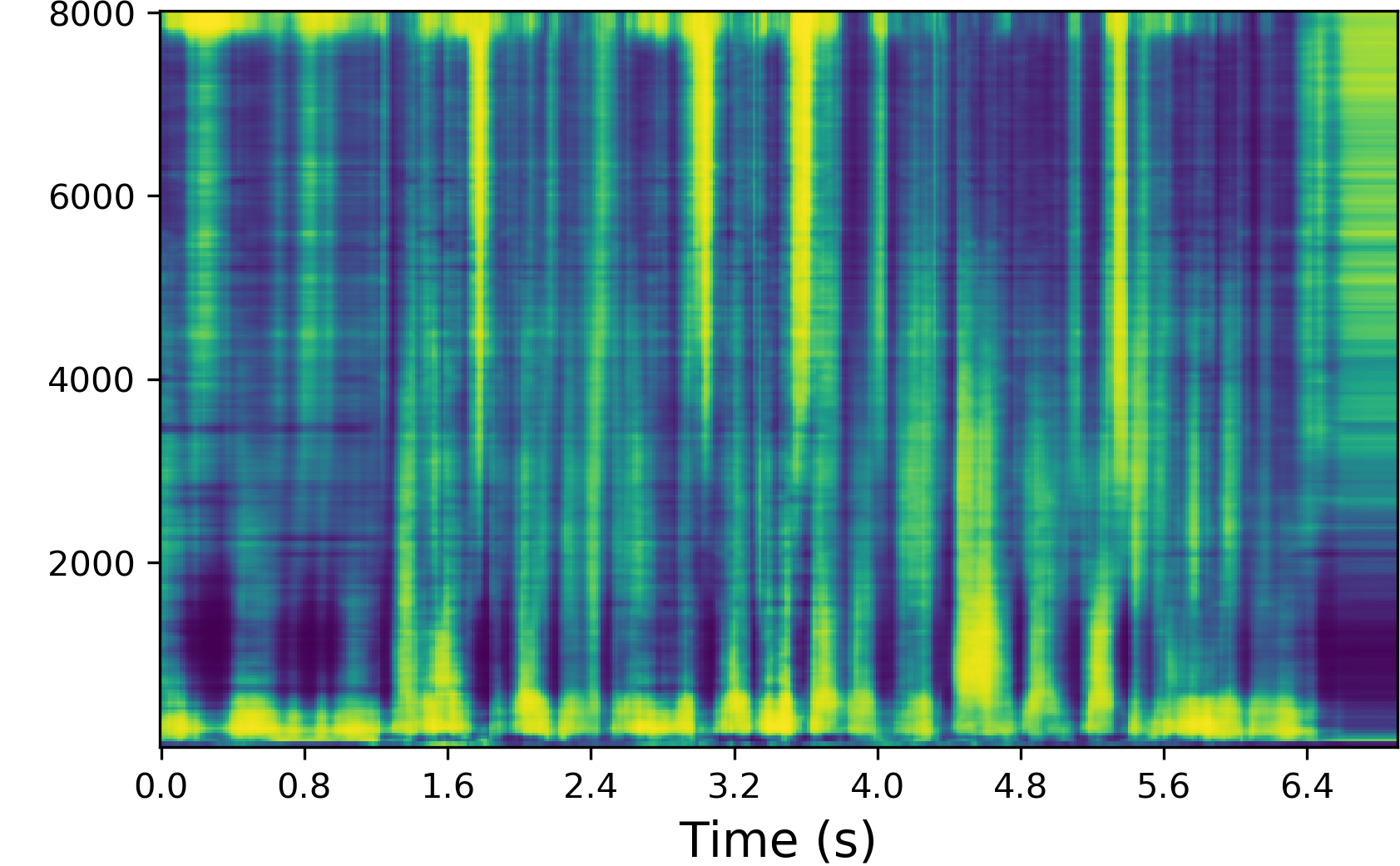}%
		\label{subfig:m1}}
	\subfloat[]{\includegraphics[width=2.4in]{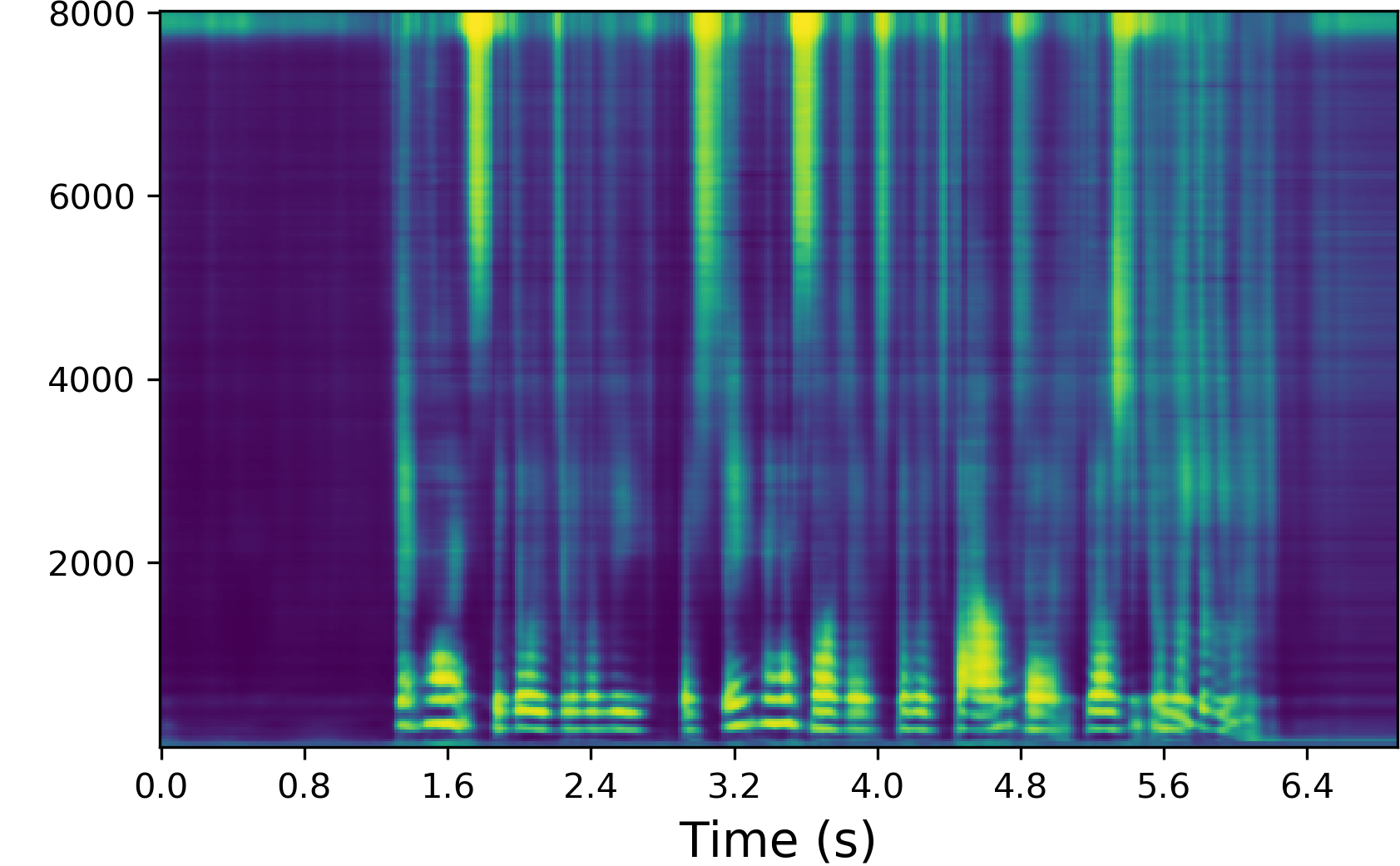}%
		\label{subfig:m2}}
	\hfil
	\caption{\ac{tf} representations of the signals at the first node of a microphone array. \protect\subref{subfig:input} Noisy input. \protect\subref{subfig:z} Compressed signal sent from node 2.
		\protect\subref{subfig:output} Output enhanced signal.
		\protect\subref{subfig:irm} Ideal ratio \ac{tf} mask corresponding to the input. \protect\subref{subfig:m1} Mask predicted by the single-node \ac{dnn}. \protect\subref{subfig:m2} Mask predicted by the multi-node \ac{dnn}.}
	\label{fig:specs}
\end{figure*}

The solution introduced in our previous work proved that using the compressed signals could help to better estimate the \ac{tf} masks, thus to increase the speech enhancement performance. We represent in Figure~\ref{fig:specs} different spectrograms throughout the processing to highlight how useful the compressed signals are in the estimation of the \ac{tf} masks. As can be seen in Figure~\ref{fig:specs}\subref{subfig:m2}, the \ac{tf} mask predicted at the second step is less noisy and more accurate than the \ac{tf} mask estimated at the first step in Figure~\ref{fig:specs}\subref{subfig:m1}, especially at lower frequencies where the different harmonics can be clearly identified. This leads to the filtered signal represented in Figure~\ref{fig:specs}\subref{subfig:output}, where a higher noise reduction can be observed.
In this paper, we propose to extend this solution to more various scenarios, and to cases where the signals sent can be either the estimation of the target signal or that of the noise, depending on the needs at the receiving node. We propose a detailed analysis of the aspects that have an impact on the final speech enhancement performance. Based on this, we improve the solution by optimizing the training of the \acp{dnn} and by taking full profit of the signals sent among nodes.

\subsection{Single-node networks}
\begin{figure}[h!]
	\centering
	\subfloat[Using the distant TF mask]{
		\includegraphics[width=\linewidth]{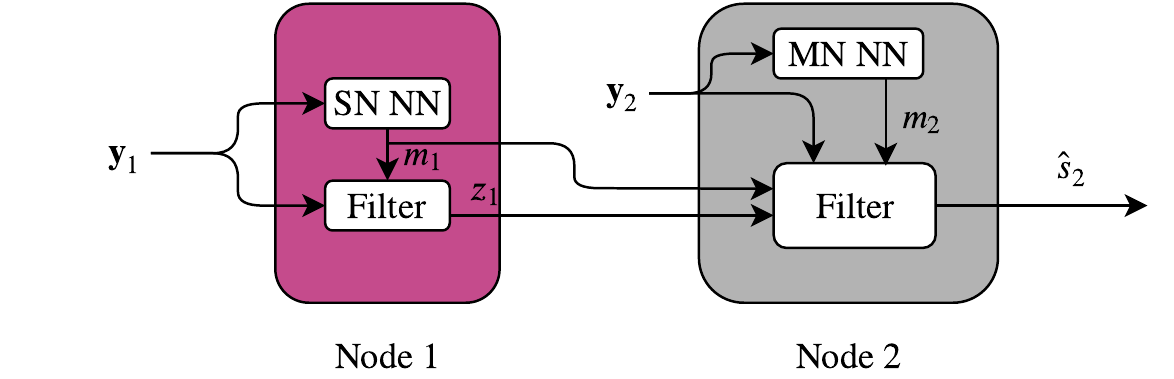}%
		\label{subfig:mask_on_z_distant}}
	\hfil
	\subfloat[Using the local TF mask]{
		\includegraphics[width=\linewidth]{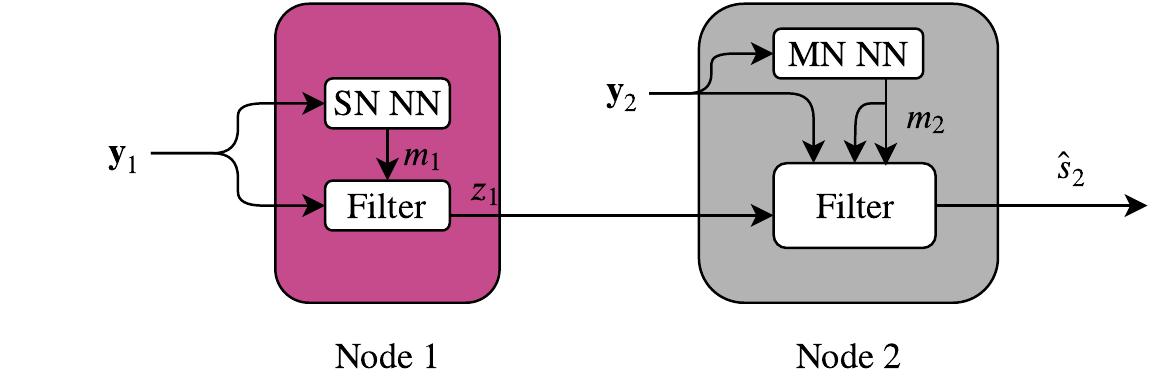}%
		\label{subfig:mask_on_z_local}}
	\caption{Representation of how using the local TF mask can spare some bandwidth cost.}
	\label{fig:mask_on_z}
\end{figure}

In their original version of \ac{danse}, Bertrand and Moonen assumed that all the nodes of the network share the same \ac{vad}. That is to say that, when estimating the global filter, at a given time frame, the same binary value was used to estimate the signal statistics for both the reference signal and the compressed signals. This relies on the hypothesis that the speech activity typical variation lasts less than a frame. In our context, as can be seen in Figure~\ref{fig:step2}, \ac{tf} masks are used and the spectral variation of the speech activity should also be considered (see Equation \eqref{eq:rss}). Since the signals $\mathbf{y}_k$ of a same device are very similar, the same \ac{tf} mask is used for all the channels of $\mathbf{y}_k$, but the \ac{tf} masks of the potentially distant nodes $j\neq k$ should be sent together with $\mathbf{z}_{-k}$ in order to compute $\mathbf{R}_{ss, k}$ and $\mathbf{R}_{nn, k}$ accurately. This is represented in Figure~\ref{fig:mask_on_z}. It translates into a bandwidth overload and we experiment whether we can spare some bandwidth costs by using the \ac{tf} mask corresponding to $\mathbf{y}_k$ instead of the \ac{tf} mask corresponding to each $z_{j,\,j\neq k}$.

In addition, we study the influence of the noise diversity in the training data, in a similar manner as Kolb\ae k et al. \cite{Kolbaek2016}, but where the effects of \ac{ssn} and real-life noises are analysed separately, so as to distinguish their respective contribution to the training efficiency. \ac{ssn} is easy to create and overlaps with speech in the \ac{stft} domain, representing a cheap but challenging interference, although it is stationary and not representative of real-life noises. On the other hand, real recordings of everyday-life noises are more realistic but require much more time to gather. Our experiment aims at exploring whether the diversity and representativeness brought by the real-life noises can help improving the performance or if training on \ac{ssn} alone would be sufficient. Likewise, as our proposed solution is evaluated on various spatial configurations, we explore the influence of the spatial configuration it is trained on. We study whether a network should be trained on a specific spatial configuration to achieve high performance on it, or if a trade-off can be found between specificity and generalizability across the spatial configurations at test time.

\subsection{Multi-node networks}
The results of our previous paper were obtained on a rather simple dataset, where the spatial configurations were not so diverse and the nodes close to each other. In this paper, as will be described in Section \ref{subsec:data}, the microphones in the array can record very different signals. An illustration of this phenomenon is given in Figure~\ref{fig:sigex}. The second part of our work starts by verifying that our previous conclusions generalize well on various scenarios. In addition, as the spatial information brought by the compressed signals might be of different interest depending on the receiving node, we propose to repeat in the multi-node context the study relative to the generalizability across spatial configurations.
Besides, we investigate the impact of low \acp{snr} on the performance of single-node and multi-node \acp{dnn}. Related to the low \ac{snr} issue, we consider the quality of the compressed signals needed to train the multi-node network. Indeed, the multi-node \ac{dnn} required at the second filtering step is trained on the compressed signals obtained at the first step on the training dataset. These compressed signals can be obtained by using either the \ac{irm} at the first filtering step of the algorithm, as in the previous section, or the \ac{tf} mask predicted by the single-node \ac{dnn}. Since the multi-node network is tested with the compressed signals resulting from a predicted \ac{tf} mask, it seemed natural to use the predicted \ac{tf} masks to train the multi-node \ac{dnn}. However, the single-node \acp{dnn} are not perfect (see e.g. the mask of Figure~\ref{fig:specs}\subref{subfig:m1}) and the resulting compressed signals might be too poor to build useful training data. That is why we check which of two multi-node \acp{dnn} performs best when one is trained with compressed signals computed with \acfp{irm} and one is trained with compressed signals computed with predicted \ac{tf} masks. In both cases, the performance is evaluated when the compressed signals are obtained with the predicted \ac{tf} mask.

\subsection{Signals sent among nodes}
\begin{figure}
	\centering
	\includegraphics[width=0.9\linewidth]{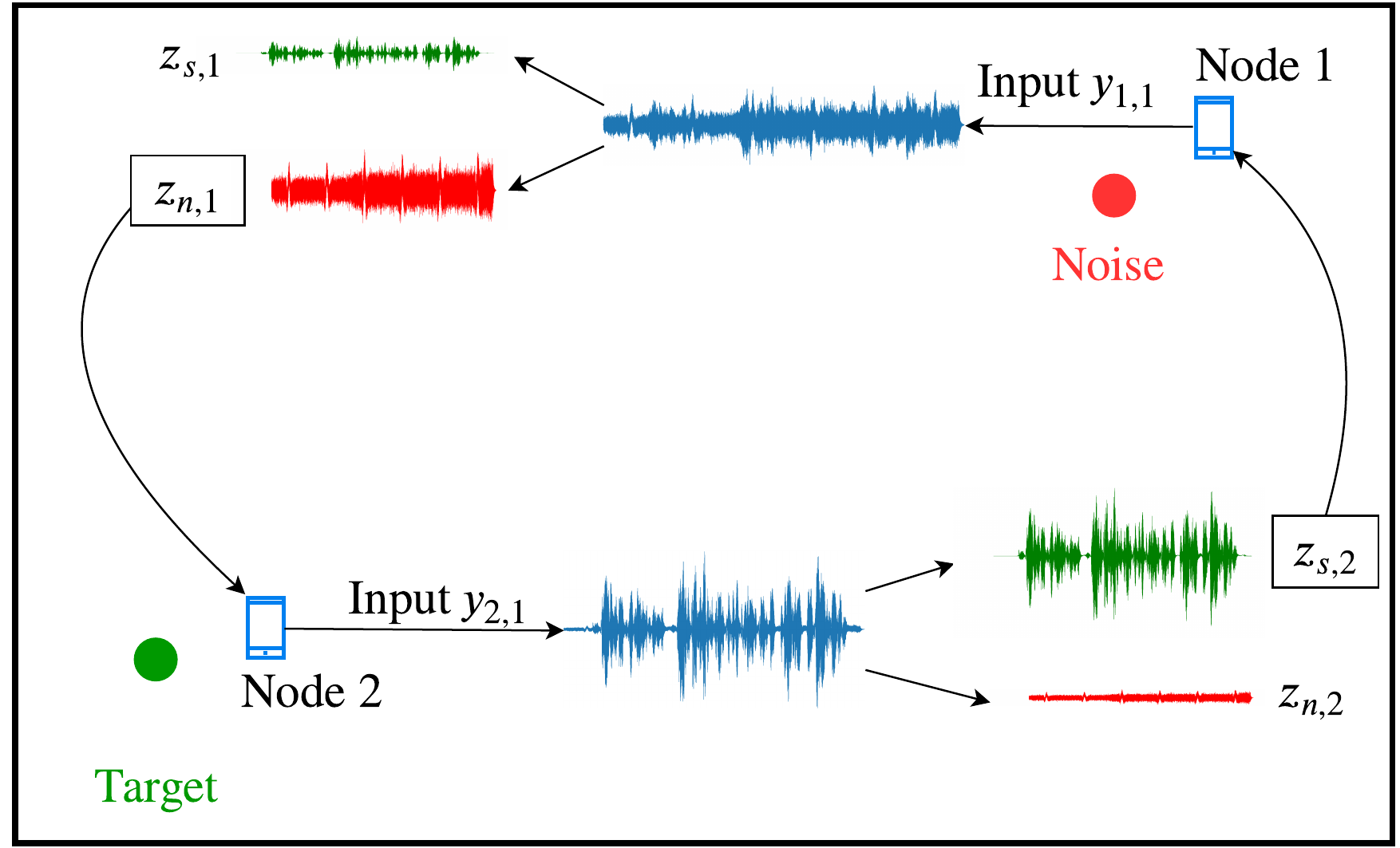}
	\caption{Example of a situation highlighting the importance of the information provided by the compressed signals. The first node, close to the noise source, can accurately estimate the noise component and send it to the second node. The second node, close to the target source, can accurately estimate the target component and send it to the first node.}
	\label{fig:sigex}
\end{figure}

In a third part, we analyse the importance of adequately selecting the compressed signals sent to the other nodes. Indeed, each node can estimate both the speech and noise components of a noisy signal. In a speech enhancement context, the target signal is the speech signal, but the noise signal may contain very useful information as well, since the \ac{mwf} also requires the estimation of the noise statistics. Both the speech and noise signals are useful to estimate the speech and noise covariance matrices, and it has also been shown that even a coarse estimation of the noise can help to increase the output performance of a \ac{dnn} for speech enhancement in the context of automatic speech recognition \cite{Perotin2018a}. An example of this phenomenon is represented in Figure~\ref{fig:sigex} where two nodes see a very different view of the same acoustic scene because of their locations in the room. 
We check which signal (i.e. the estimation of the noise or the estimation of the target speech) a given node should send to the other nodes depending on its location in the room.

\section{Setup}
\label{sec:setup}
\subsection{Datasets}
\label{subsec:data}
\begin{figure*}[!t]
	\centering
	\subfloat[\textit{Random room} configuration]{
		\includegraphics[width=.3\linewidth]{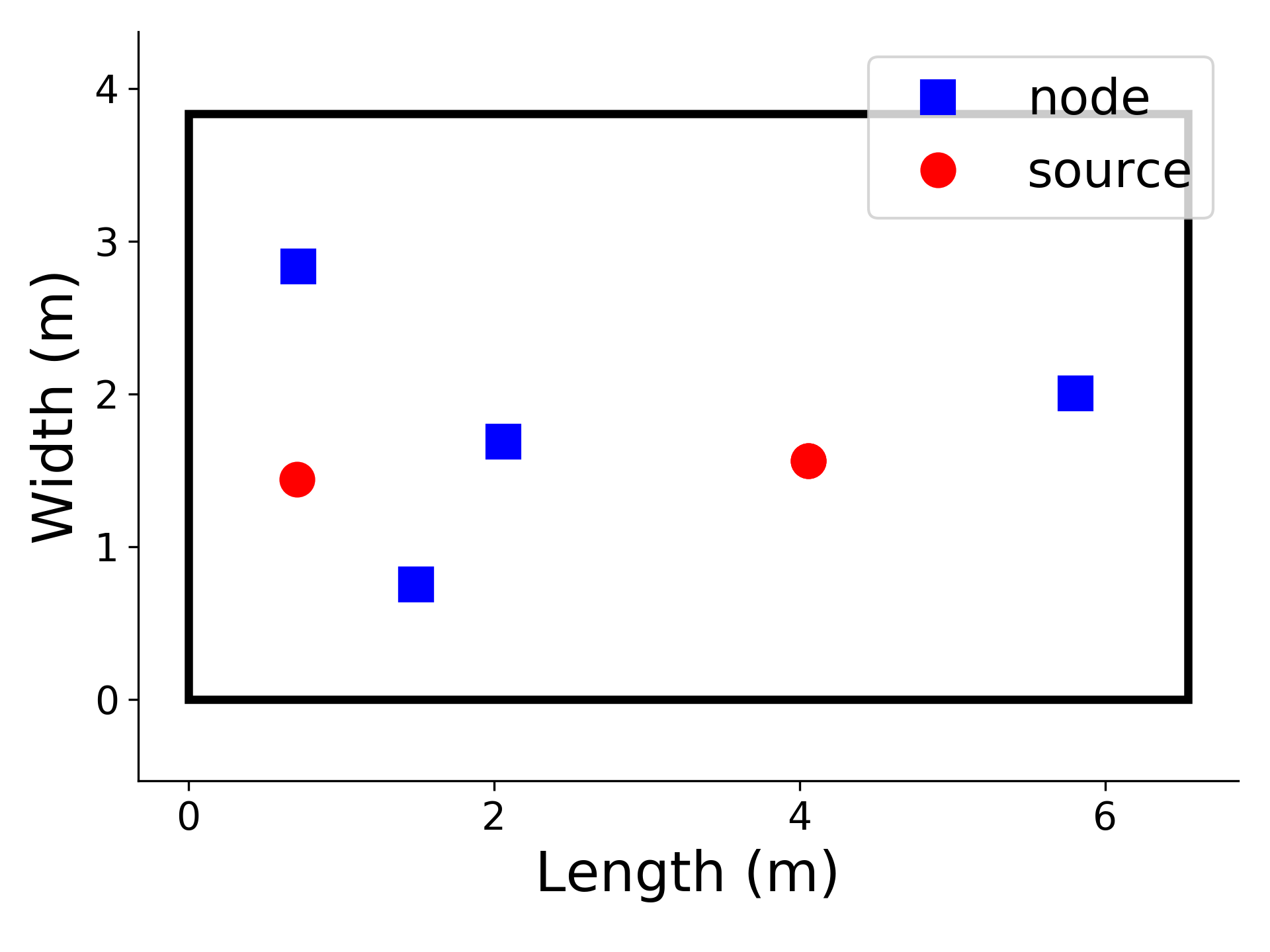}%
		\label{subfig:random}}
	\subfloat[\textit{Living room} configuration]{
		\includegraphics[width=.3\linewidth]{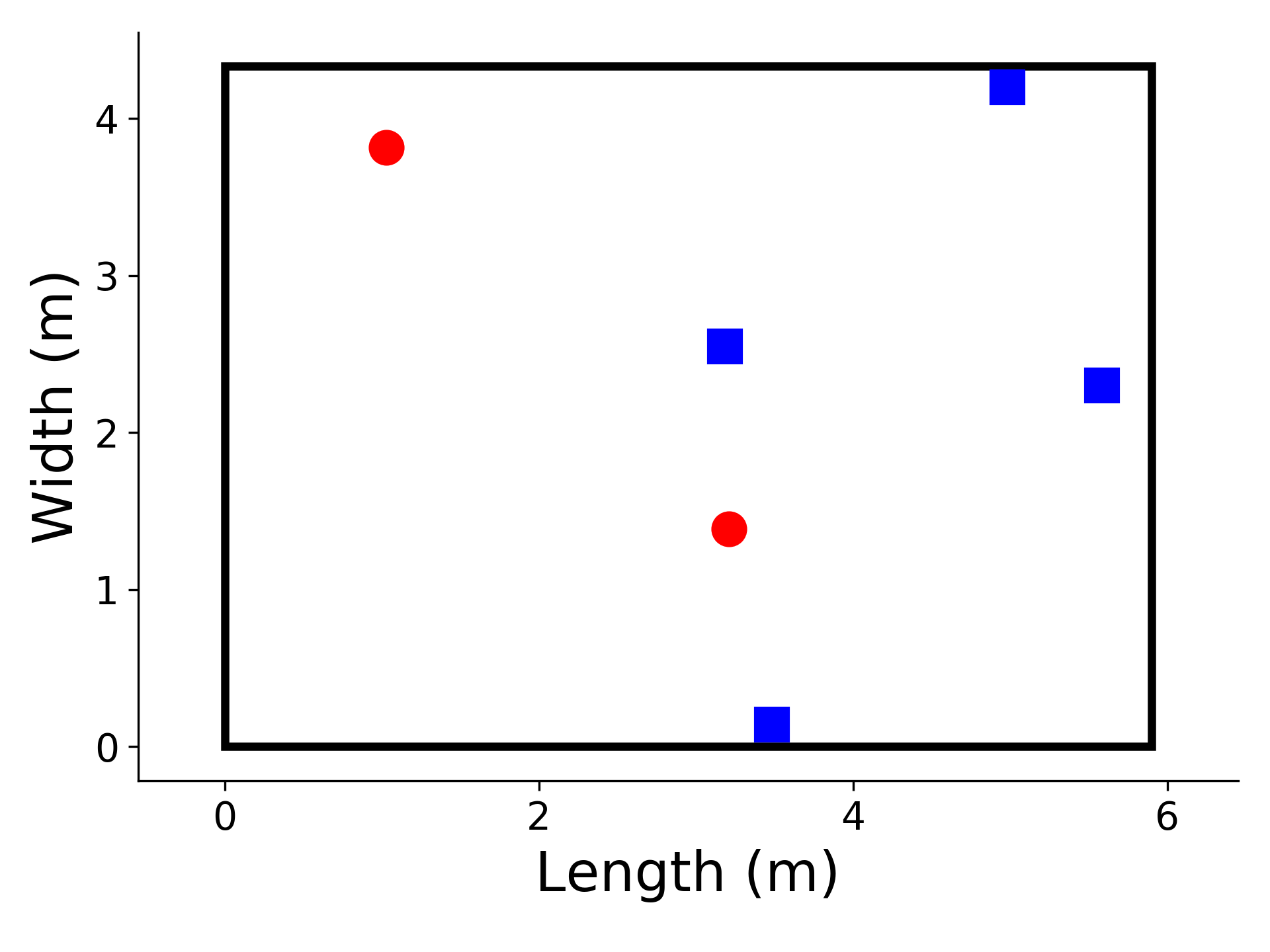}%
		\label{subfig:living}}
	\subfloat[\textit{Meeting room} configuration]{
		\includegraphics[width=.3\linewidth]{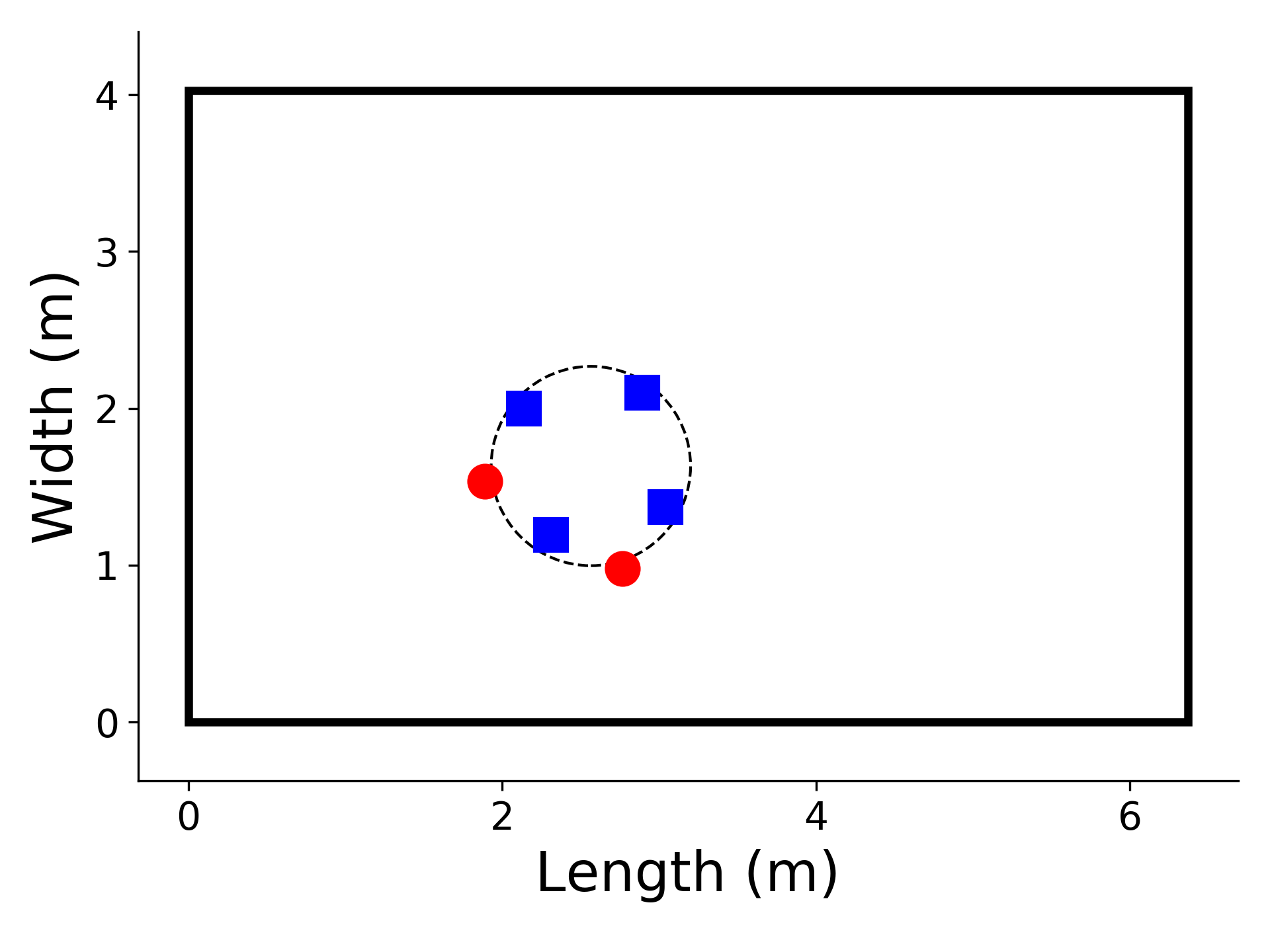}%
		\label{subfig:meeting}}
	\hfil
	\caption{2D representations of the three spatial configurations. The acoustic effect of the table (dashed-line circle in \protect\subref{subfig:meeting}) is not simulated; it is only represented for a better visualization.}
	\label{fig:configs}
\end{figure*}
We create three spatial scenarios with the Python toolbox Pyroomacoustics \cite{Scheibler2018}. An example of each scenario can be seen in Figure~\ref{fig:configs}. Two of these datasets, called \textit{living room} and \textit{meeting room}, aim at simulating the real-life scenarios that correspond to two typical use cases of a living room and a meeting room. To see whether training the \acp{dnn} on one generic dataset could generalize well on the test sets of the \textit{living room} and \textit{meeting room}, we create the \textit{random room} configuration, which is less constrained and covers the specific cases of the \textit{living room} and \textit{meeting room}. 

In each scenario, shoebox-like rooms are created with a reverberation time randomly selected between 0.3~s and 0.6~s, the length between 3~m and 8~m, the width between 3~m and 5~m and the height between 2.5~m and 3~m. $K = 4$ recording devices (called \textit{nodes} in the rest of the paper) are simulated, each embedded with four microphones ($M_k = 4 \quad \forall k\in\llbracket 1; K\rrbracket$). The microphones are at a distance of 5~cm to the node center. Two sources, one target source and one noise source, are added.
In each scenario, the speech content is taken from the LibriSpeech clean subsets \cite{Panayotov2015}. The noise source can be either \ac{ssn} or a real recording of everyday-life noises, downloaded from Freesound \cite{Freesound}\footnote{The noise dataset is available at \url{https://zenodo.org/record/4019030}.}. The noise source signals are amplified by a random gain between -6~dB and 0~dB. After convolution, most of the \acp{snr} between lie in the range [-10; +10]~dB depending on the node position in the room.

The first scenario, called \textit{random room} (see Figure~\ref{fig:configs}\subref{subfig:random}), has very few additional constraints. The two sources and the nodes are randomly placed in the room with the only constraints that they all should be distant of at least 50~cm from each other and from the walls. The nodes are at a random height between 0.7~m and 2~m, as if they were recording devices laid on a piece of furniture, or hearings aids worn by an impaired person. The sources are between 1.20~m and 2~m high, to fit the standard height of most noise sources.

The second scenario, called \textit{living room} (see Figure~\ref{fig:configs}\subref{subfig:living}) recreates a situation that could typically happen in a living room with one target speech source and one interference noise source. Three nodes are placed within 50~cm from the walls as if they were on shelves and the fourth device is placed randomly in the room, at 50~cm at least of the walls and the other nodes. All the nodes are at a random height between 0.7~m and 0.95~m. The two sources are also randomly placed in the room at 50~cm  at least from the nodes and the walls and at a random height between 1.20~m and 2~m.

The third scenario, called \textit{meeting room} (see Figure~\ref{fig:configs}\subref{subfig:meeting}) simulates a meeting configuration where two people are sitting around a table. One speaker is the target speaker while the second one is considered as an interferent source. The table is circular, with its radius randomly chosen between 0.5~m and 1~m, its height randomly chosen between 0.7~m and 0.8~m and its center randomly placed in the room. The nodes are placed every 90$^\circ$ on the table, at a random distance between 5~cm and 20~cm from the table edge. The two sources are randomly placed around the table within 50~cm from the table edge, at a random height between 1.15~m and 1.3~m, and at 15~cm at least from the walls. The reflection of the table is not simulated.

Each dataset is split into a training set containing 10000 samples of 10~s each, a validation set containing 1000 samples of 10~s and a test set containing 1000 samples whose duration range from 6~s to 10~s. The test dataset does not overlap with the training and validation sets in terms of LibriSpeech speakers and Freesound users\footnote{A Python implementation of the code that enabled us to create these datasets is available at \url{https://github.com/nfurnon/disco}}.

\subsection{Experimental settings}
All the signals are sampled at 16000~Hz. The \ac{stft} is computed using a Hanning window of 32~ms with an overlap of 16~ms.
The same \ac{crnn} architecture is used for all experiments. The convolutional part is made of three convolutional layers with 32, 64 and 64 filters respectively, with kernel size $3 \times 3$ and stride $1 \times 1$. Each convolutional layer is followed by a batch normalisation and a maximum-pooling layer of kernel size $4 \times 1$ (no pooling over the time axis). The recurrent layer is a 256-unit GRU, followed by a fully-connected layer with a sigmoid activation function in order to map the output of the network between 0 and 1. The network was trained with the RMSprop optimizer \cite{rmsprop}. The input of the model are \ac{stft} windows of 21 frames and the ground truth targetted are the corresponding frames of the \ac{irm}.

\subsection{Performance evaluation}
\subsubsection{Metrics}
In the following, all the performance measures are quantified based on the \ac{sir} and \ac{sar}, computed with the \textit{mir\_eval}\footnote{\url{https://github.com/craffel/mir_eval/}} toolbox. These metrics require a reference signal and it was shown that they are very sensitive to the chosen reference \cite{LeRoux2019, Drude2019}. Both the source (non-reverberated) and image (reverberated) signals are valid references and quantify differently the performance. Considering the source signal as the reference enables one to keep a constant reference for all sensors despite the diversity of what they capture. However, is does not allow us to distinguish the distortion due to reverberation from the distortion due to the filter. On the other hand, considering the image signals as references enables to quantify the effects of the proposed filters only, but the implicit reference of the \ac{gevd} filter at a specific node might not be the explicit reference channel of the metric (see Section V in \cite{Serizel2014}). To cope with this, we quantify the speech enhancement performance with three metrics. The first metric is the difference between the output \ac{sir} and the input \ac{sir}\footnote{Since we do not simulate any microphone noise, the input \ac{sir} is equal to the input \ac{snr}.} when the image signals are taken as references\footnote{We noticed that the \ac{sir} was quite consistent across the reference signals, whether they were the source signals or the image signals.}. It is denoted by $\Delta SIR_{\text{cnv}}$ where the subscript $_{\text{cnv}}$ means that the reference signals are the convolved signals. The second metric is the \ac{sar} where the clean (target and noise) signals captured by a sensor of the node are considered as the references. We arbitrarily take the first microphone of each node as the reference of this node. This metric is denoted as $SAR_{\text{cnv}}$. The third metric is the \ac{sar} where the source signals are the references. It is denoted as $SAR_{\text{dry}}$ where the subscript $_{\text{dry}}$ means that the reference signals are the source signals. The difference between $SAR_{\text{dry}}$ and $SAR_{\text{cnv}}$ could be interpreted as the distortion due to the reverberation of the source signals. By keeping both of these metrics, we can quantify both the problems of denoising and that of dereverberation.

\subsubsection{Signals considered for the evaluation}
Depending on the context, we might be interested in having one well-estimated target signal for the whole microphone array, or one well-estimated target signal for each node of the microphone array. In most cases, one signal would be enough for the whole array, but it might require to send this signal to all the other nodes, resulting in a possibly undesired bandwidth overload. The question of a node-specific speech enhancement algorithm issue has also been discussed by Markovich-Golan et \textit{al}. \cite{Markovich-Golan2015}. In our case, we will mainly focus on estimating the best possible signal for the whole array, this is why, unless mentioned otherwise, the results presented in the remainder of the paper represent the average over the whole test set of the performance at the best output node, \textit{i.e.} at the node with the highest output \ac{sir}. However, we will also analyse more in detail the behaviour of the proposed solution at the node with the highest and lowest input \ac{sir} in Sections \ref{subsec:low_sir} and \ref{sec:zs_zn}, in order to highlight the cooperation among nodes in the microphone array and the needs of the nodes concerning the compressed signals that they receive. This will then be mentioned explicitly.

\section{Analysis of the performance with single-node networks}
\label{sec:optim_sn}
This section focuses on several factors that impact the performance of the proposed algorithm using masks estimated with single-node \acp{dnn}. As described in Section \ref{subsec:mask_based_se}, the \ac{danse} algorithm is split in two steps and the compressed signals are sent between nodes to compute the filter of the second step, but the same single-node network is used for both steps.

\subsection{Importance of node-specific \ac{tf} masks}
\label{subsec:loc_vs_dist}
In this section, we investigate in oracle conditions which \ac{tf} mask should be applied on the compressed signal. To do so, we compare two cases. In the first case, the \ac{tf} mask of the node sending the signal (called \textit{distant} node) is applied to the compressed signal in order to compute the speech and noise statistics at the second filtering step. In the second step, the \ac{tf} mask of the receiving node (called \textit{local} node) is used. The results are reported in Table \ref{tab:maskz} where the two cases are respectively referred to as \textit{distant} and \textit{local}.

\begin{table}
	\centering
	\caption{Speech enhancement performance in oracle conditions in the random room configuration when applying the distant or local oracle \ac{tf} mask on the compressed signal. The best significant results are in bold.}
	\begin{tabular}{|l|c|c|c|}
		\hline
		(dB) & $\boldsymbol{\Delta SIR_{\text{cnv}}}$ & $\boldsymbol{SAR_{\text{cnv}}}$ & $\boldsymbol{SAR_{\text{dry}}}$ \\
		\hline
		local & 26.8 $\pm$ 0.4 & \textbf{10.9 $\mathbf{\pm}$  0.2} & \textbf{9.6 $\mathbf{\pm}$  0.2} \\
		\hline
		distant & 26.1 $\pm$  0.4 & 8.3 $\pm$  0.2 & 9.0 $\pm$  0.2 \\
		\hline
	\end{tabular}
	\label{tab:maskz}
\end{table}

As can be seen in Table \ref{tab:maskz}, using the \ac{tf} mask of the local node instead of the distant node not only limits the bandwidth requirements, but also increases the speech enhancement performance in terms of \ac{sar} without decreasing the \ac{sir}. This might come from the fact that the beamformer is robust to small \ac{tf} mask estimation errors. The drop of $SAR_{\text{cnv}}$ when the distant \ac{tf} mask is used could be due to the fact that the filtered signal is closer to the reference of the distant nodes than to the reference of the local node. This could decrease the metric without actually decreasing the performance. The almost equal $SAR_{\text{dry}}$ between the two metrics seems to  confirm this hypothesis. As a conclusion, in the remainder of the paper, the local \ac{tf} mask will be the one applied on all the compressed signals coming from the other nodes to estimate the signal statistics required by the \ac{mwf}.

\subsection{Robustness to unseen noise}
\label{subsec:noise}
\begin{figure}
	\centering
	\hspace{-0.4cm}
	\subfloat[Results on test set with speech shaped noise]{
		\includegraphics[width=.5\linewidth]{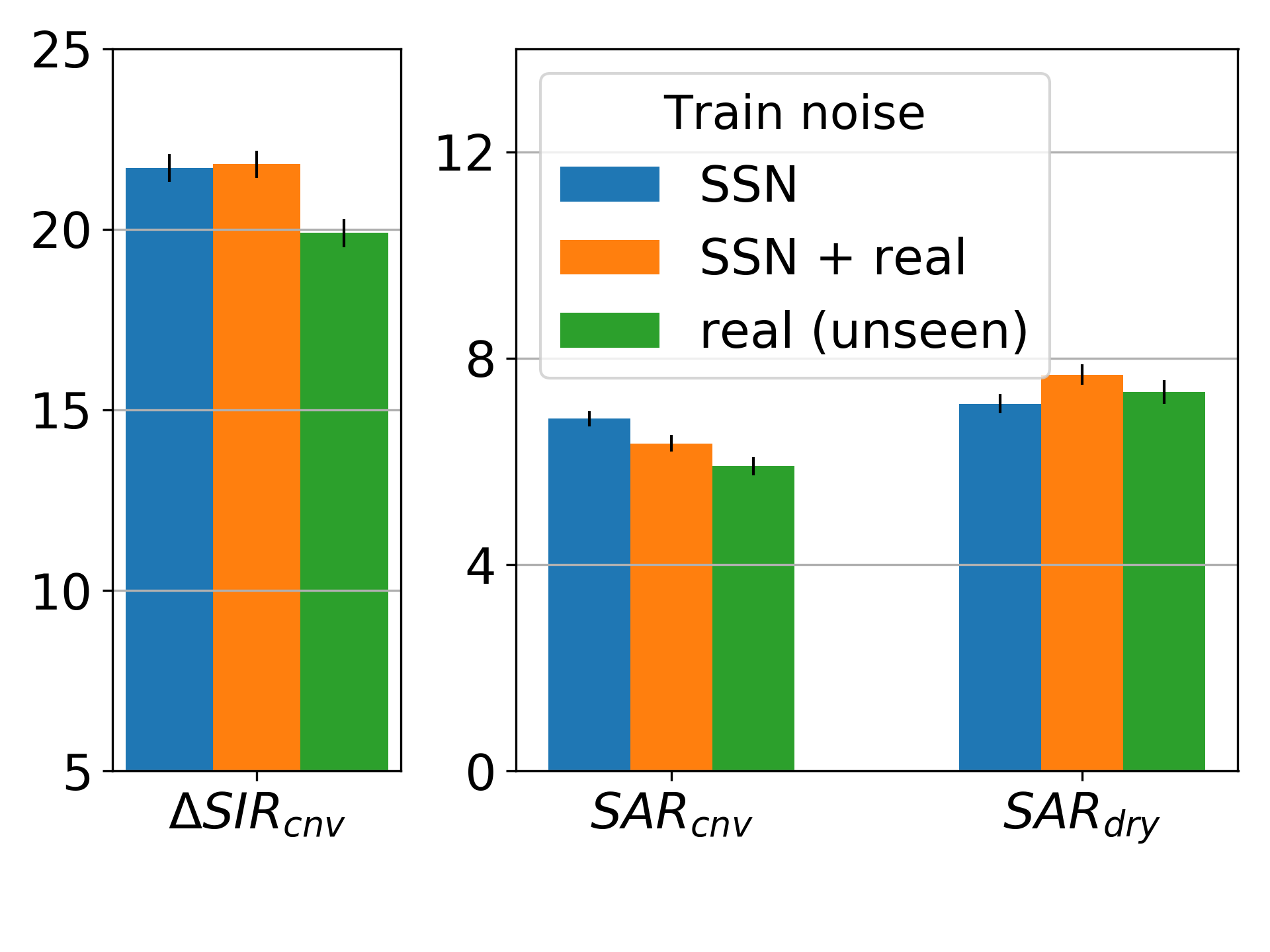}%
		\label{subfig:noise_ssn}}
	\subfloat[Results on test set with real noise]{
		\includegraphics[width=.5\linewidth]{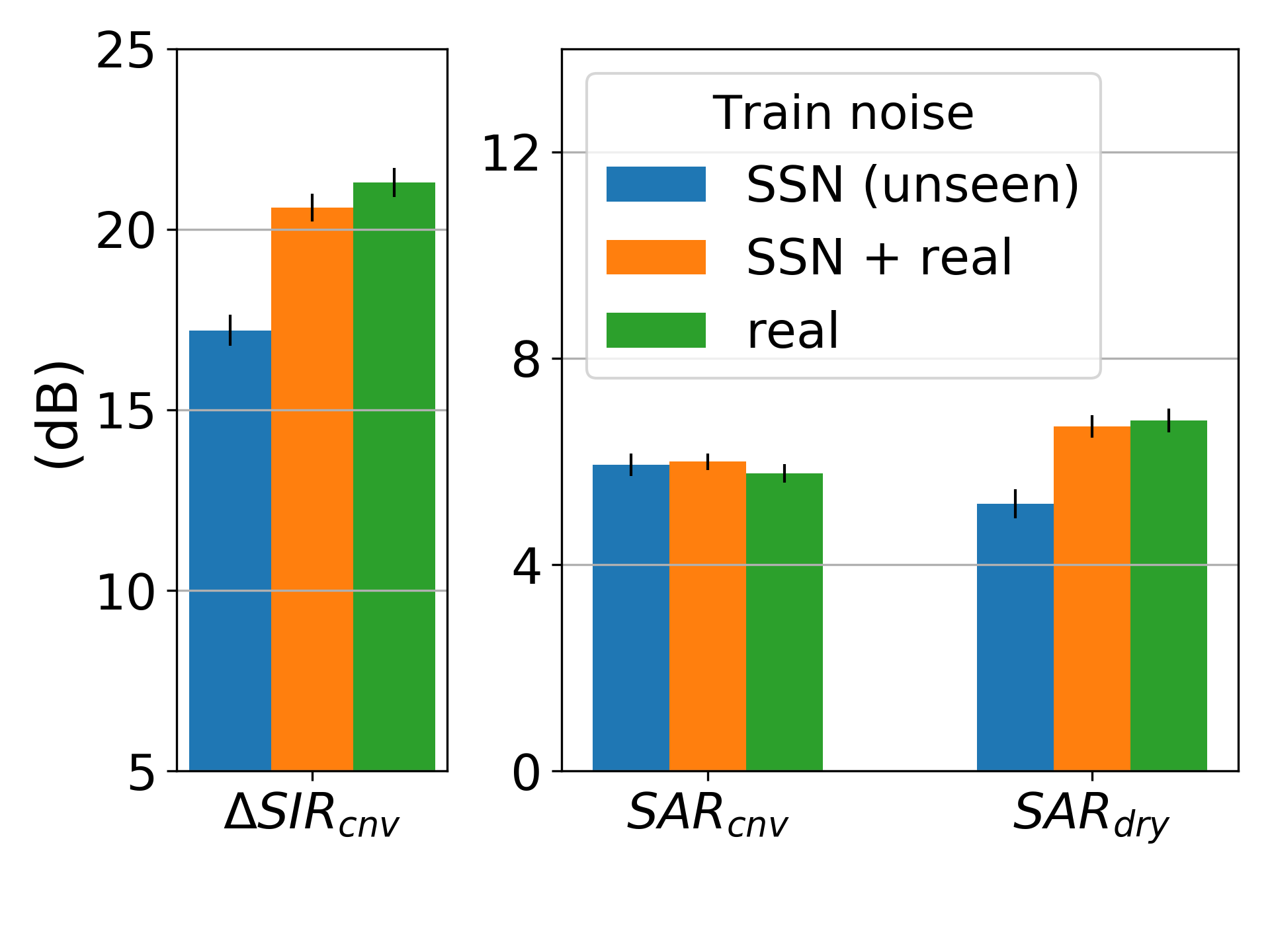}%
		\label{subfig:noise_fs}}
	\caption{Speech enhancement performance of the single-node DNNs in the random room configuration for different training and test noise conditions. The $\Delta SIR_{\text{cnv}}$ difference in Figure~\ref{subfig:noise_fs} between the two last networks is not statistically significant.}
	\label{fig:noise}
\end{figure}

We trained a model in the random room configuration under three noise conditions. In the first condition, the noise signals are all samples of \ac{ssn}. In the second condition, the noises are real recordings of everyday-life noises as described in Section \ref{sec:setup}. In the third condition, the model is trained with half of the signals mixed with \ac{ssn} noise and the other half mixed with real noises. The three resulting models are tested on noisy signals where the noise is either \ac{ssn} or a real recording. The corresponding results are represented in Figure~\ref{fig:noise}, where \textit{real} refers to recordings downloaded from Freesound.

The first observation is that the networks trained on a single type of noise are specialized on this noise, i.e. they perform better in matched test conditions than in unseen conditions. This is especially true in terms of $\Delta SIR_{\text{cnv}}$. On the other hand, the network trained on both types of noises performs at least as well as the specialized network. This conclusion is similar to the conclusion of Kolb\ae k et al \cite{Kolbaek2016}. However, because we separately analysed the influence of the \ac{ssn} and of the real noise, our experiment is additionally able to show that removing the \ac{ssn} from the training set decreases the generalization capacities of the \ac{dnn}, in particular on stationary noises.

As a conclusion to this section, a wider variety of training material leads to a robust network that performs as good as a specialized network in matched conditions, and can maintain performance in unmatched conditions. In the following, since the test set might contain unseen noises during the training, all the networks will be trained on both types of noises described above in order to increase their robustness, but they will be tested on the real noises.

\subsection{Robustness to an unseen spatial configurations}
\label{subsec:spat_sn}
We now consider the impact of the spatial scenario while training the \ac{dnn}. We compare three \acp{dnn}, trained on the signals generated in the three spatial configurations introduced in Section \ref{subsec:data}, and tested on each of these scenarios.
\begin{figure*}[!t]
	\centering
	\subfloat[Random room]{
		\includegraphics[width=.3\linewidth]{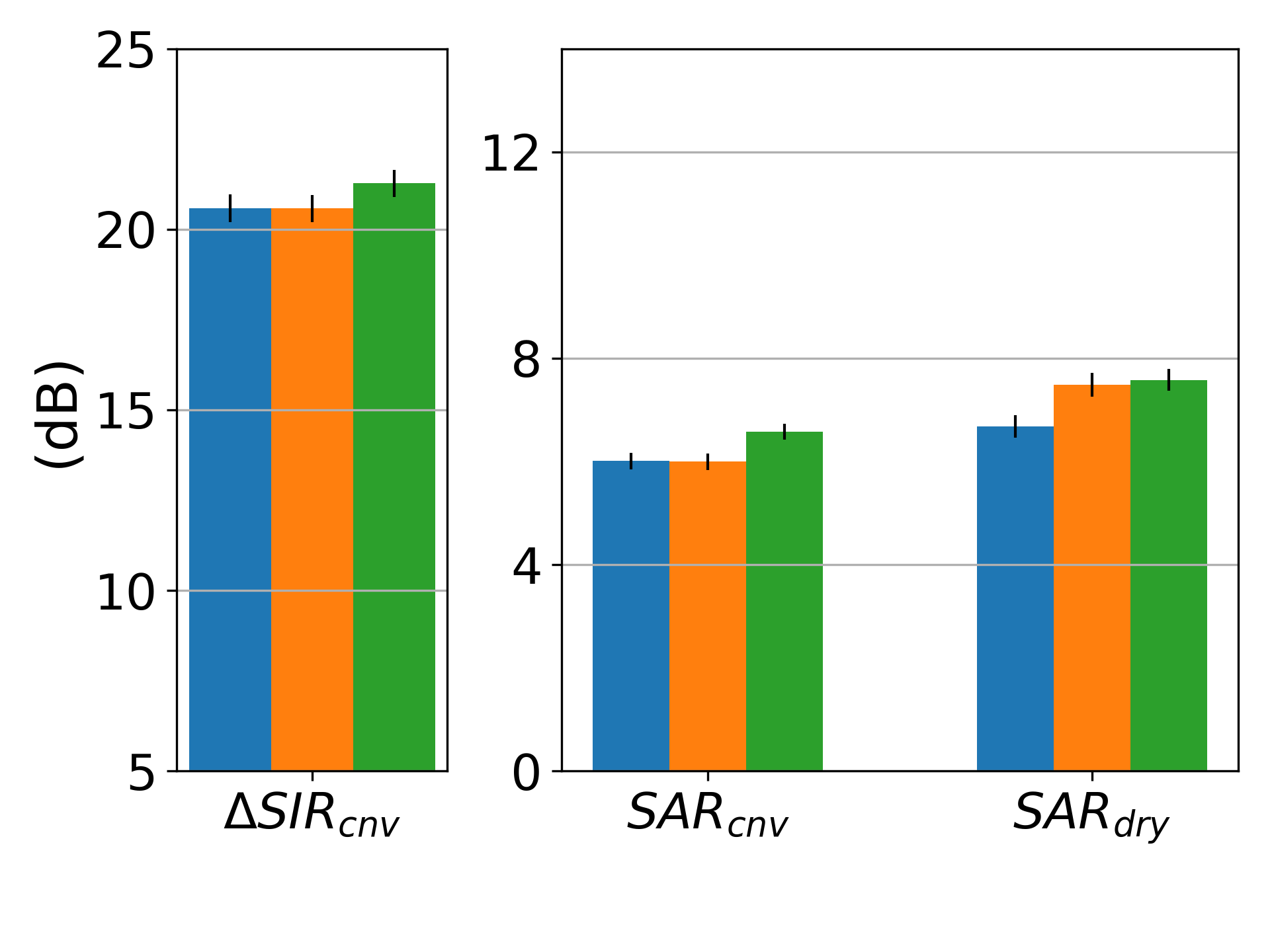}%
		\label{subfig:cc_random}}
	\subfloat[Living room]{
		\includegraphics[width=.3\linewidth]{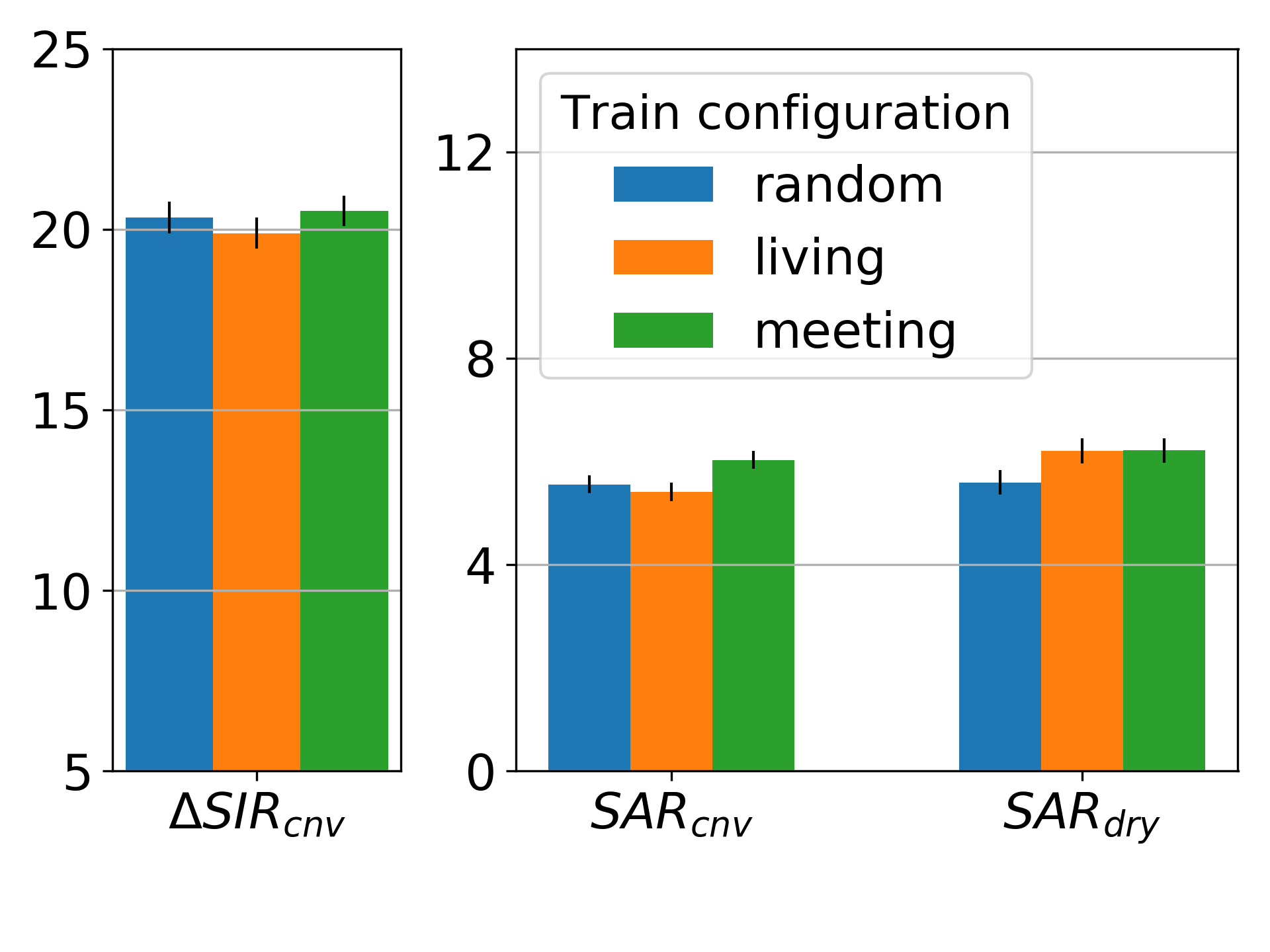}%
		\label{subfig:cc_living}}
	\subfloat[Meeting room]{
		\includegraphics[width=.3\linewidth]{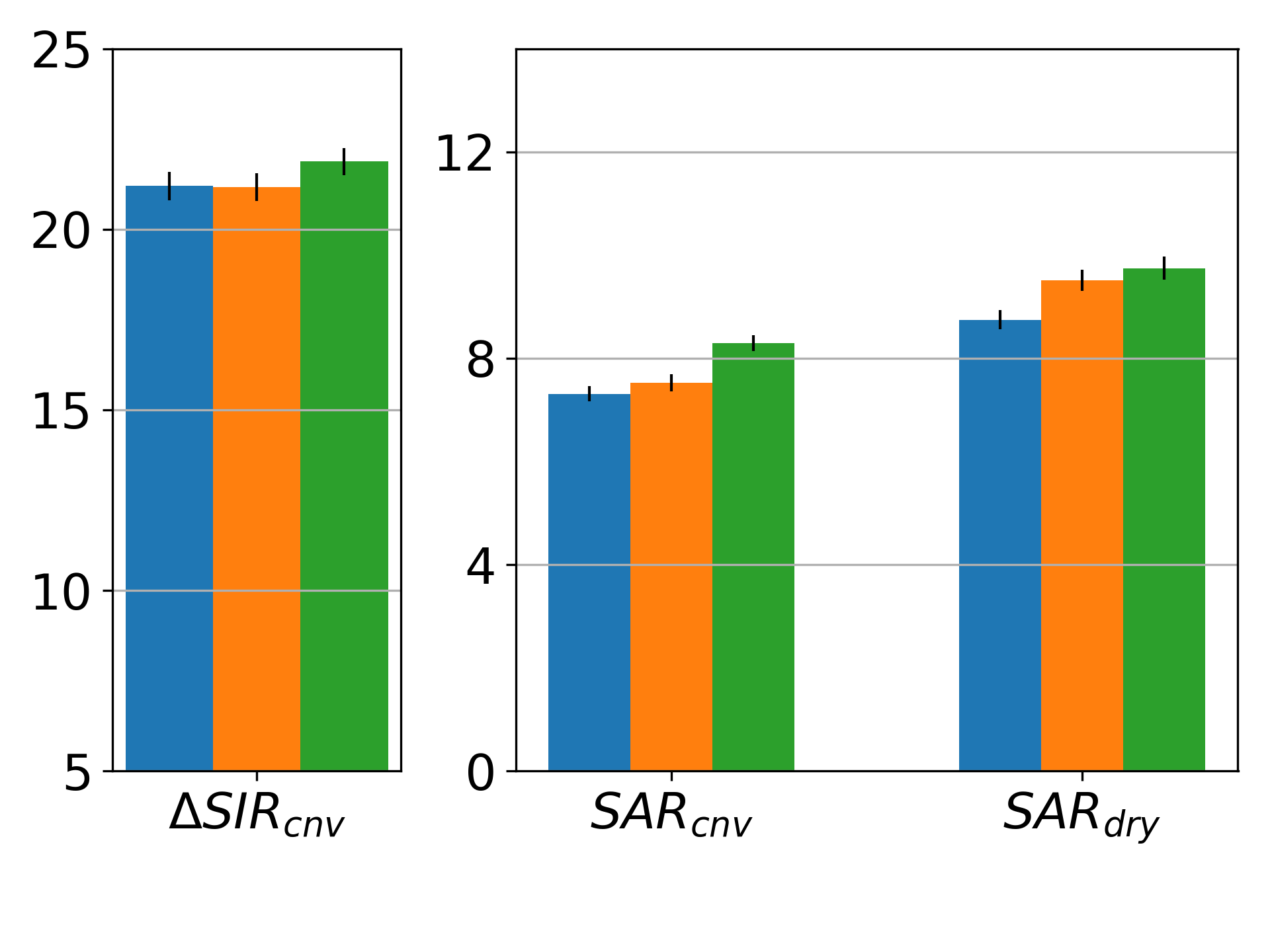}%
		\label{subfig:cc_meeting}}
	\hfil
	\caption{Speech enhancement performance of three single-node networks trained each on a different spatial configuration.}
	\label{fig:spat_sn}
\end{figure*}

As can be seen in Figure~\ref{fig:spat_sn}, only mildly significant differences can be observed between the three models. One exception can be highlighted, when the \ac{dnn} trained on the \textit{meeting room} configuration yields the best results, probably because, due to the closeness of some nodes to the noise source, this \ac{dnn} has seen more challenging scenarios during the training, making it more robust. Apart from this specific single-node scenario, it would not have a big impact to train on one spatial configuration and test on another.

In particular, it is interesting to notice that the $SAR_{\text{dry}}$ values are higher in the \textit{meeting room} configuration than in the two other ones. This is because the microphones are close to the target source, which is hence less distorted by the reverberation. This confirms the relevance of the third metric.

\section{Analysis of the performance with the multi-node networks}
\label{sec:optim_mn}
We now extend our study to the case where, at the second filtering step, the \ac{dnn} also receives the signals from the other nodes and uses them as additional input to predict the \ac{tf} masks. In a similar manner to our previous work, the signals sent are all estimations of the target signal \cite{Furnon2020}.

\subsection{Benefit of using multi-node DNNs}
\label{subsec:sn_vs_mn}
\begin{figure*}[!t]
	\centering
	\subfloat[Random room]{
		\includegraphics[width=.3\linewidth]{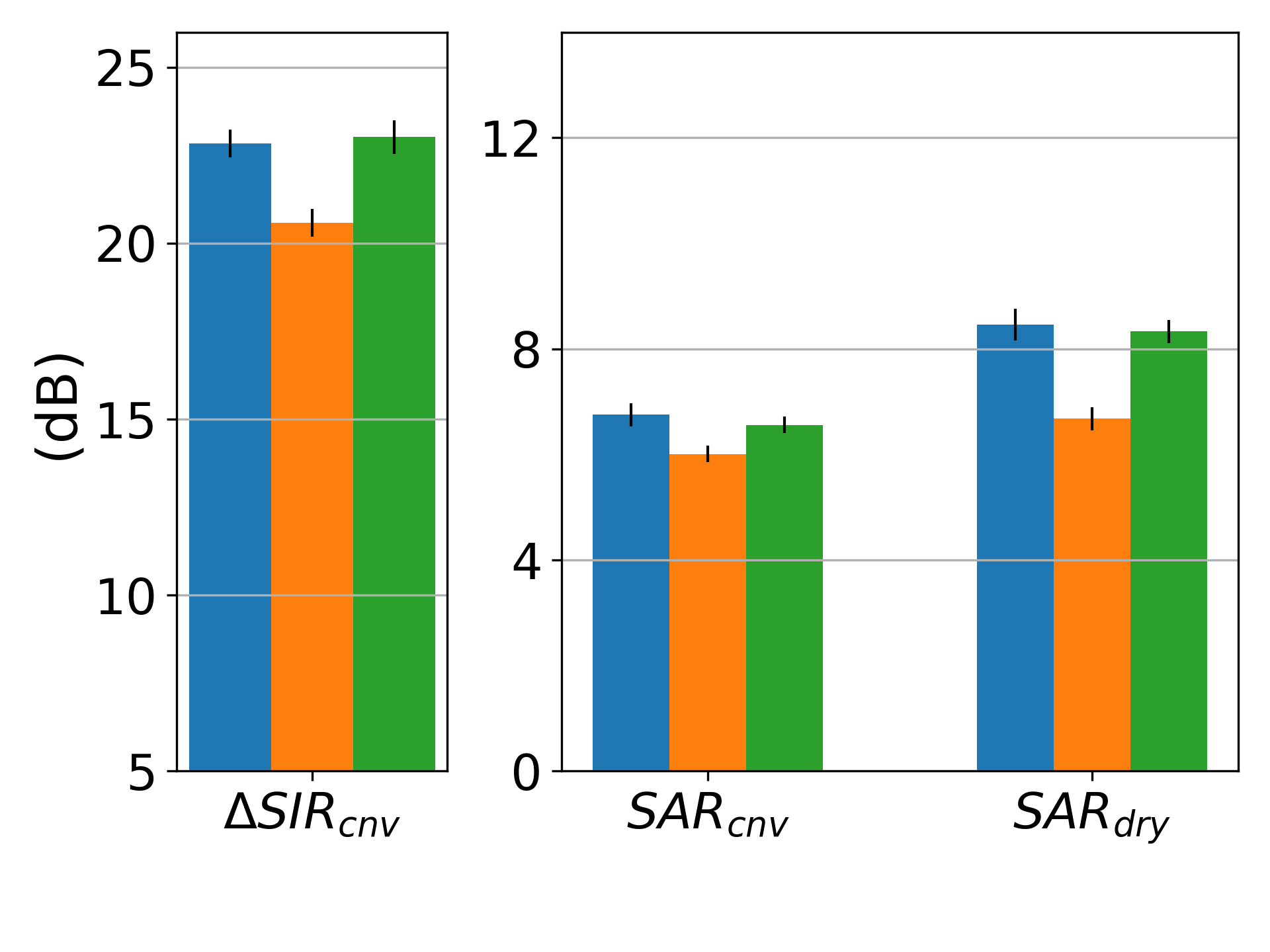}%
		\label{subfig:sn_vs_mn_ran}}
	\subfloat[Living room]{
		\includegraphics[width=.3\linewidth]{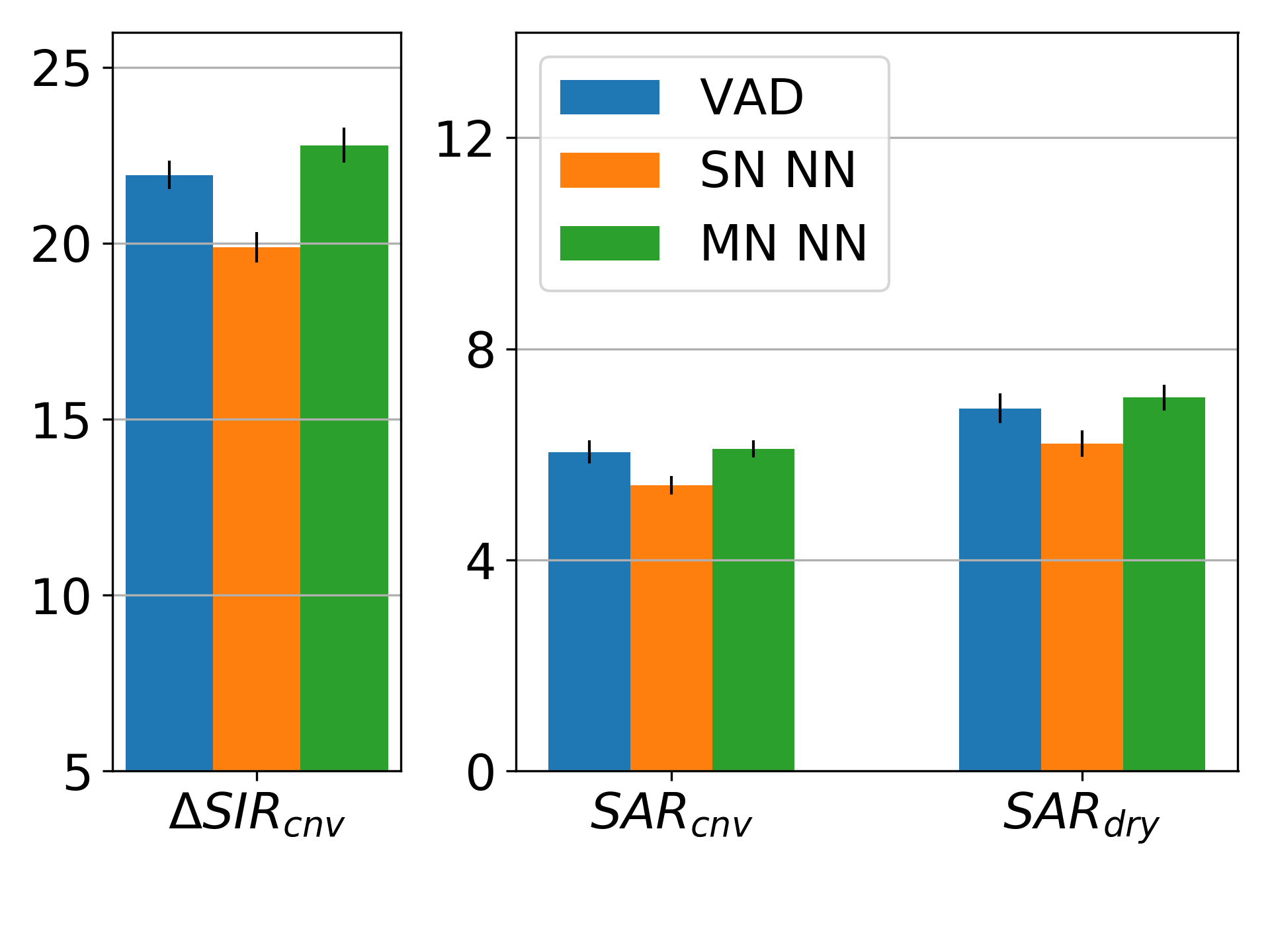}%
		\label{subfig:sn_vs_mn_liv}}
	\subfloat[Meeting room]{
		\includegraphics[width=.3\linewidth]{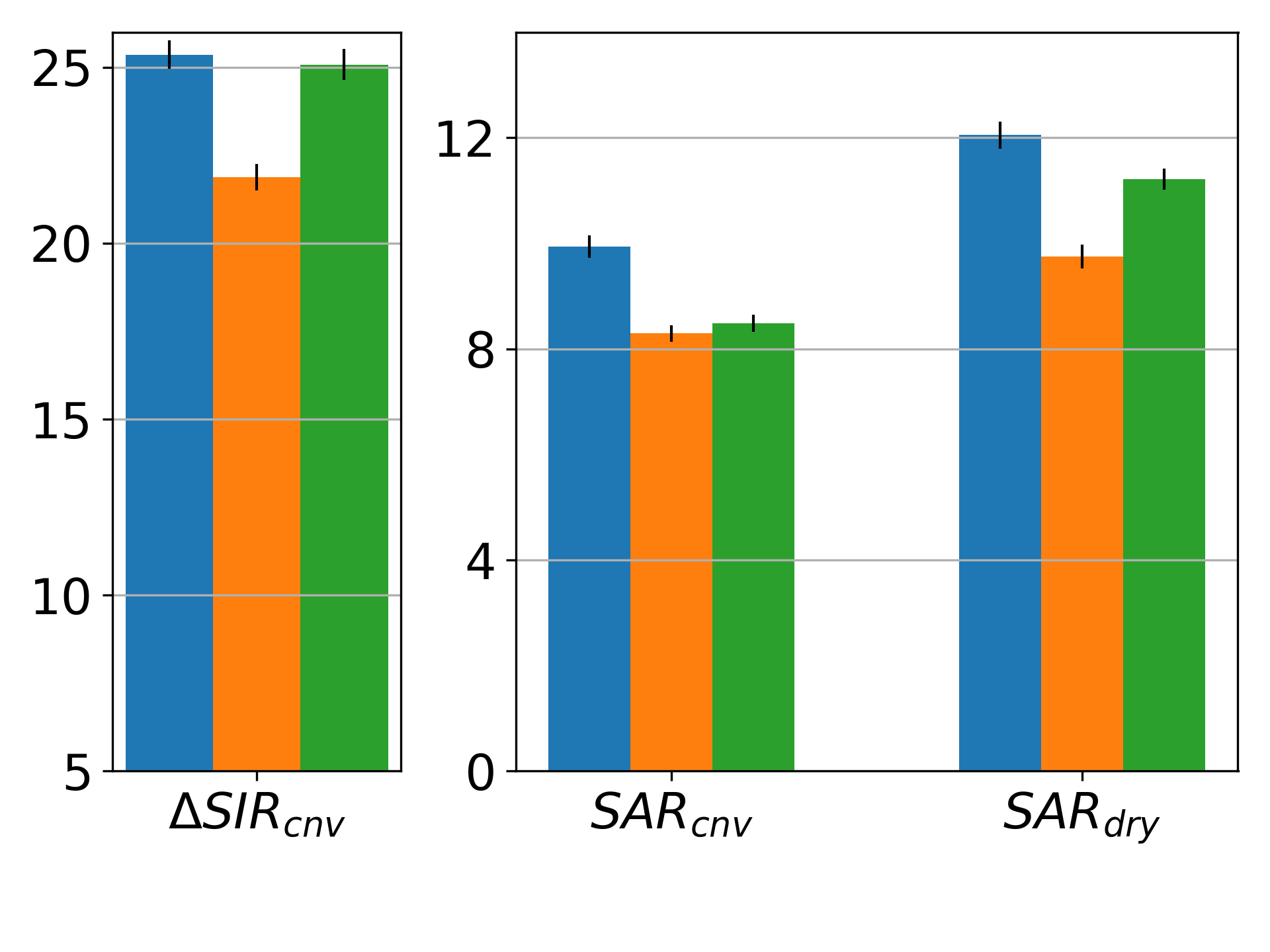}%
		\label{subfig:sn_vs_mn_mee}}
	\hfil
	\caption{Speech enhancement performance on the three spatial configurations using an oracle \ac{vad}, a single-node \ac{dnn} and a multi-node \ac{dnn} at the second filtering step.}
	\label{fig:sn_vs_mn}
\end{figure*}

We first study in this section the advantage of using multi-node \acp{dnn} over single-node \acp{dnn} at the second filtering step. We train a multi-node \ac{dnn} on each of the spatial configurations introduced in Section \ref{subsec:data}. The compressed signals used to train the \ac{dnn} are obtained with \acp{irm}, but at test time, the \acp{irm} are replaced by the \ac{tf} masks estimated by the single-node \acp{dnn}. The results are given in Figure~\ref{fig:sn_vs_mn} and compared to the oracle case of \ac{danse}, where the signals statistics are computed with an oracle \ac{vad}.

The multi-node \ac{dnn} brings an $\Delta SIR_{\text{cnv}}$ improvement of around 3~dB compared to the single-node \ac{dnn} (see Figure~\ref{fig:spat_sn}), and up to 1.5~dB improvement in terms of $SAR$. Besides, it increases the performance up to what can be achieved with an oracle \ac{vad} in terms of $SAR$, except on the \textit{meeting room} configuration where the scenario is more challenging. The conclusion of our previous paper, saying that the compressed signals are useful to better predict the \ac{tf} masks, is thus confirmed on three real-life scenarios.

\subsection{Influence of the spatial configuration}
\label{subsec:spat_mn}
Given that the compressed signals convey a lot of spatial information, the conclusions of Section \ref{subsec:spat_sn} might not hold in the multi-node \ac{dnn}. We repeated the experiments in the multi-node case, where the compressed signals are given at the input of the \acp{dnn}. 
Similarly to the conclusions to Section \ref{subsec:spat_sn}, there was very little difference across the three \acp{dnn}. 
That is why we will consider only one network in the sequel, the one trained and tested on the \textit{random room}.

\subsection{Exploitation of the mixture diversity in low SIR conditions}
\label{subsec:low_sir}
\begin{figure}
	\centering
	\includegraphics[width=\linewidth]{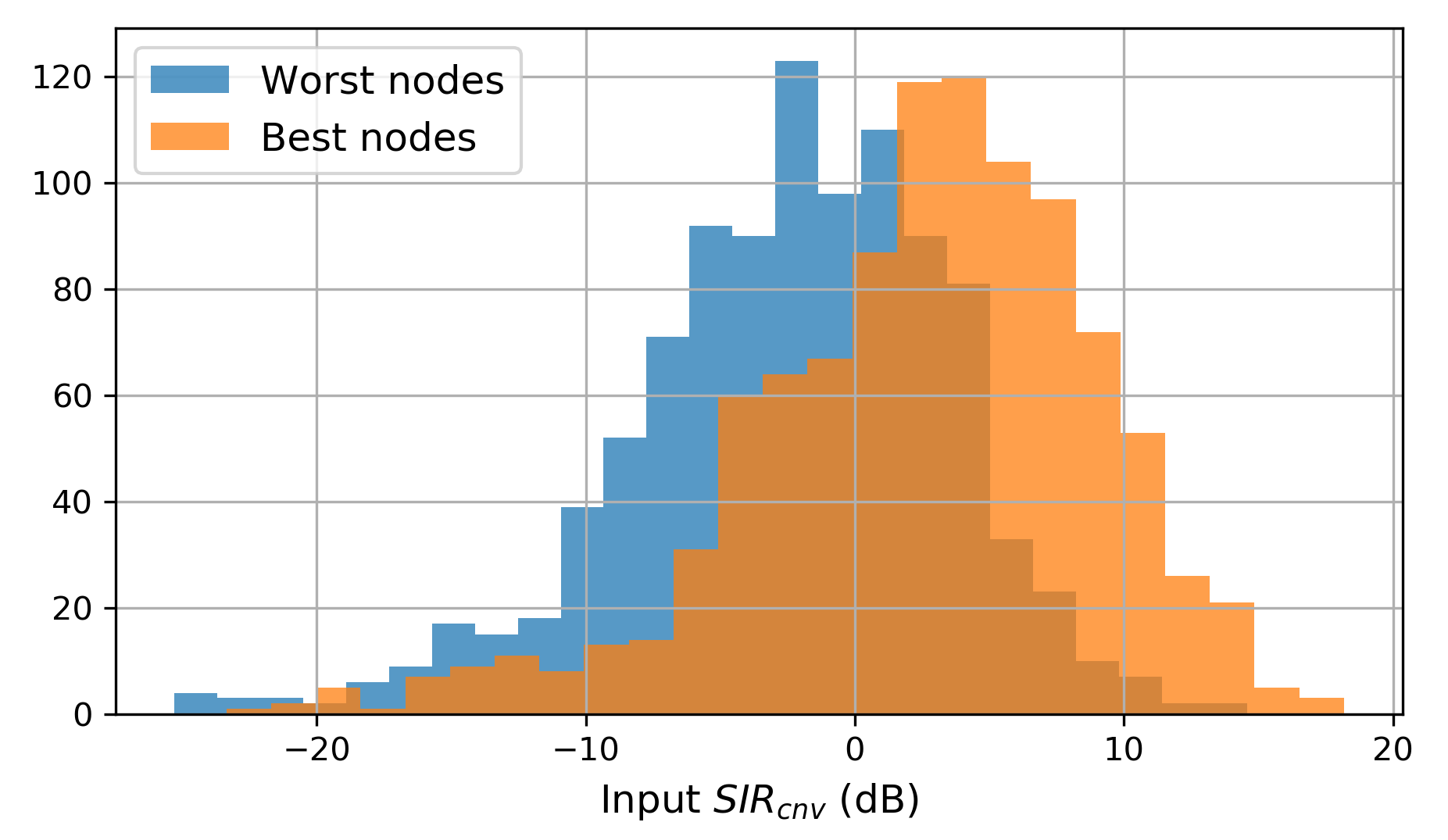}
	\caption{Histogram of the input $SIR_{\text{cnv}}$ at the best input nodes and at the worst input nodes.}
	\label{fig:bi_vs_wi}
\end{figure}

In this section, we show the advantage of using distributed microphone arrays and we highlight the cooperation among nodes of the microphone array with our multi-node solution. We report the performance of our solution at the best input node and at the worst input node of the microphone array in the \textit{random room} configuration. The best (resp. worst) input node is the node with the highest (resp. lowest) input $SIR_{\text{cnv}}$. We report these results in Table \ref{tab:sn_vs_mn} for the single-node solution (indicated by "SN") and for the multi-node solution (indicated by "MN"). To recall, in the single-node solution, the \ac{dnn} does not have the compressed signals to predict the \ac{tf} mask at any of both filtering steps. In the multi-node solution, the \ac{dnn} of the second filtering step has the compressed signals and the local noisy signal to predict the \ac{tf} mask.
The best (resp. worst) input node is indicated with the subscript $_{\text{bi}}$ (resp. $_{\text{wi}}$) in the table. The distribution of the input $SIR_{\text{cnv}}$ corresponding to the best and worst input nodes is represented in Figure~\ref{fig:bi_vs_wi}.

\begin{table}
	\centering
	\caption{Speech enhancement performance of the single-node and multi-node networks at the best and worst input nodes of the \textit{random room} configuration.}
	\begin{tabular}{|l|c|c|c|c|}
		\hline
		(dB) & $\boldsymbol{SIR_{\text{cnv}}}$ & $\boldsymbol{\Delta SIR_{\text{cnv}}}$ & $\boldsymbol{SAR_{\text{cnv}}}$ & $\boldsymbol{SAR_{\text{dry}}}$ \\
		\hline
		SN$_{\text{bi}}$ & 18.7 $\pm$ 0.6 & 16.1 $\pm$ 0.5 & 5.8 $\pm$  0.2 & 5.8 $\pm$  0.3 \\
		SN$_{\text{wi}}$ & 14.2 $\pm$ 0.7 & 16.6 $\pm$ 0.6 & 3.4 $\pm$  0.2 & 3.6 $\pm$  0.3 \\
		\hline
		MN$_{\text{bi}}$ & 20.5 $\pm$ 0.7 & 17.9 $\pm$  0.6 & 6.4 $\pm$  0.2 & 7.4 $\pm$  0.3 \\
		MN$_{\text{wi}}$ & 18.1 $\pm$ 0.7 & 20.5 $\pm$ 0.6 & 4.2 $\pm$  0.2 & 5.8 $\pm$  0.3 \\
		\hline
	\end{tabular}
	\label{tab:sn_vs_mn}
\end{table}

As can be seen in Table~\ref{tab:sn_vs_mn}, even in the single-node case, the performance is relatively good at the best input node. The $\Delta SIR_{\text{cnv}}$ is similar to the one at the worst input node, but the output $SIR_{\text{cnv}}$ is higher. Using multi-node \acp{dnn} at this best input node does improve the final performance, but in a lesser extend than at the worst input node. This is especially true when considering the $\Delta SIR_{\text{cnv}}$ which increases of almost 4~dB at the worst input node but only 1.8~dB at the best input node. This reduces considerably the discrepancy of output $SIR_{\text{cnv}}$ across the whole microphone array. It shows that the nodes cooperate and that the \ac{dnn} on the worst input node is able to exploit the information coming from the other nodes. At the worst input node, the benefit of our method is twofold: the compressed signals come from nodes with a higher \ac{sir} and additionally, they are already filtered with a well-predicted \ac{tf} mask.

As a counterpart, this also probably means that the compressed signals sent by the worst input nodes are not so useful. However, these nodes are the closest to the noise source, so the network could predict quite well the \ac{tf} mask corresponding to the noise source. Sending the noise estimation as the compressed signals could improve the overall performance. This is what we propose to analyse in Section \ref{sec:zs_zn}.

\subsection{Using oracle or predicted compressed signals}
\label{subsec:clean_train_data}
In order to know which masks should be used to compute the compressed signals of the training dataset, we compare two \acp{dnn}. The first \ac{dnn} is trained on compressed signals output by a filter with an oracle \ac{tf} mask (referred to as "oracle") and the second \ac{dnn} is trained on compressed signals output by a filter with a predicted \ac{tf} mask (referred to as "predicted"). The results are reported in Table \ref{tab:clean_train_data}.

\begin{table}
	\centering
	\caption{Speech enhancement performance when oracle masks or predicted masks are used to compute the compressed signals needed to train the multi-node \acp{dnn}. The significantly best results are indicated in bold.}
	\begin{tabular}{|l|c|c|c|}
		\hline
		(dB) & $\boldsymbol{\Delta SIR_{\text{cnv}}}$ & $\boldsymbol{SAR_{\text{cnv}}}$ & $\boldsymbol{SAR_{\text{dry}}}$ \\
		\hline
		oracle & 23.0 $\pm$ 0.5 & 6.6 $\pm$  0.2 & \textbf{8.3} $\mathbf{\pm}$  \textbf{0.2} \\
		\hline
		predicted & 23.4 $\pm$  0.5 & 6.6 $\pm$  0.2 & 7.5 $\pm$  0.2 \\
		\hline
	\end{tabular}
	\label{tab:clean_train_data}
\end{table}

The only significant difference is observed in terms of $SAR_{\text{dry}}$, where using the compressed signals computed with oracle \ac{tf} masks outperforms the method using the predicted \ac{tf} masks. This is probably explained by the fact that the training material is clean. In the rest of the paper, the results are obtained when the multi-node \acp{dnn} are trained with oracle compressed signals.

\section{Exchanging signals between nodes}
\label{sec:zs_zn}
\begin{figure*}[!t]
	\centering
	\subfloat[Best output node]{
		\includegraphics[width=.3\linewidth]{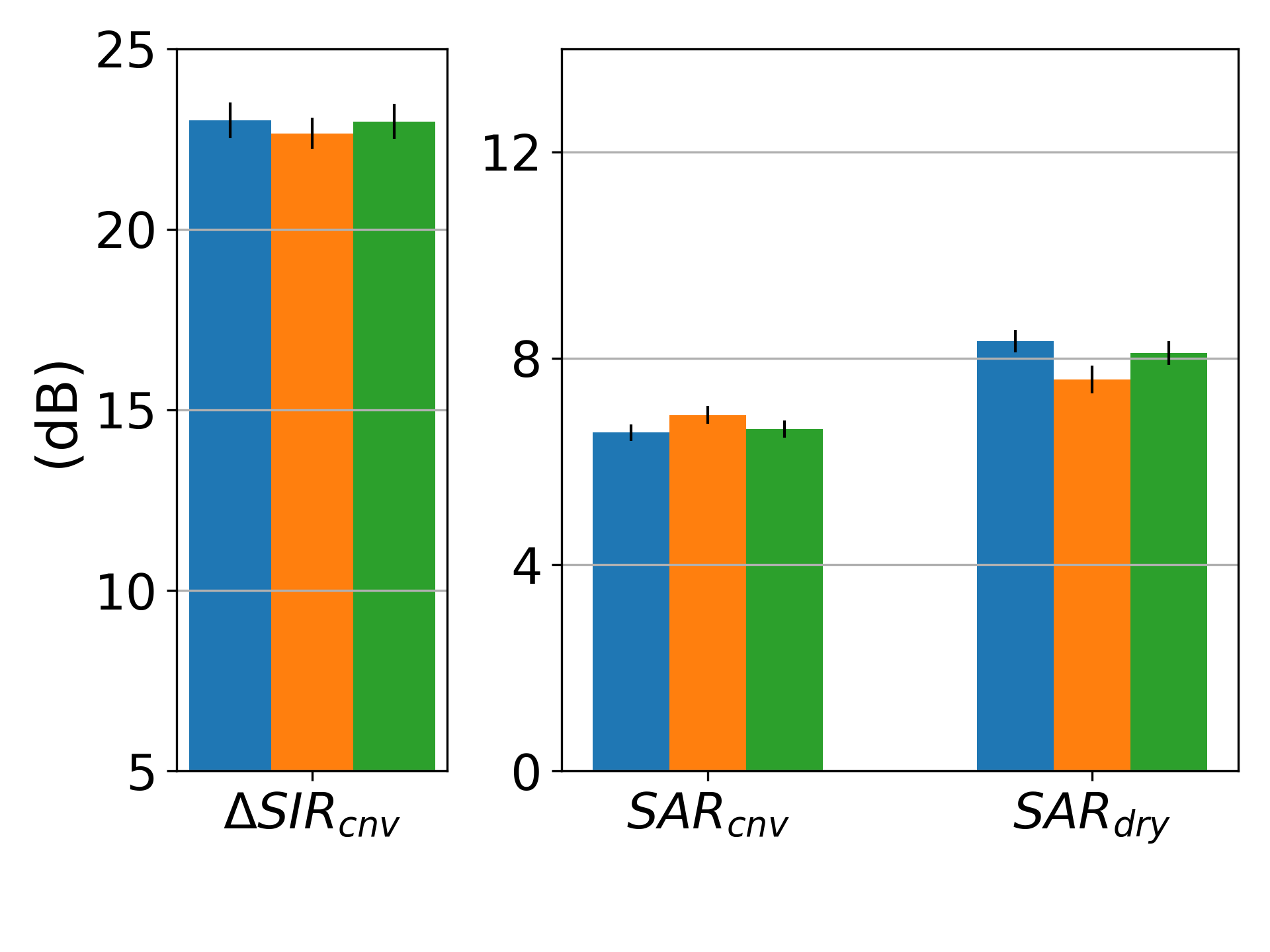}%
		\label{subfig:mn_bo_random}}
	\subfloat[Best input node]{
		\includegraphics[width=.3\linewidth]{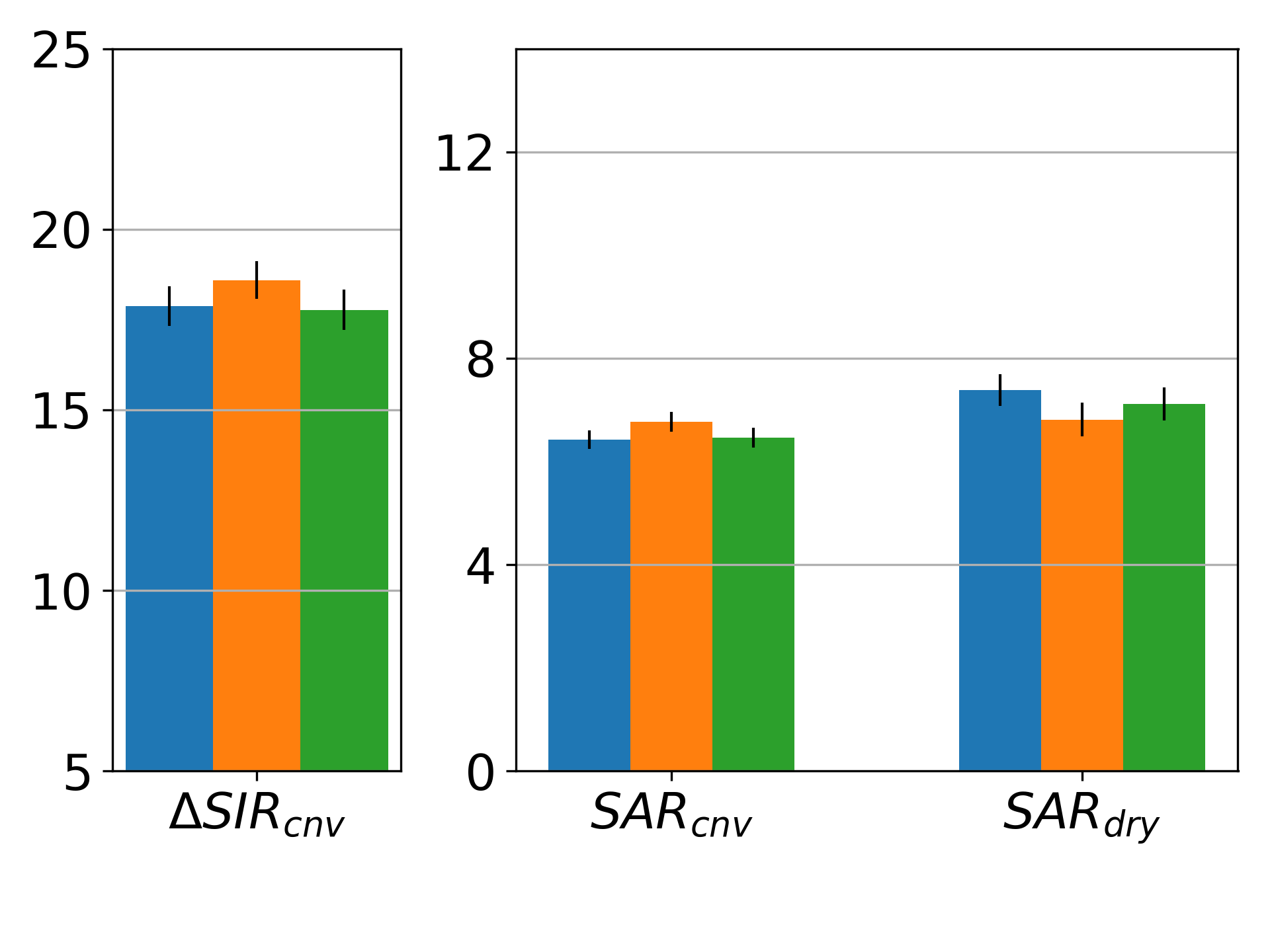}%
		\label{subfig:mn_bi_random}}
	\subfloat[Worst input node]{
		\includegraphics[width=.3\linewidth]{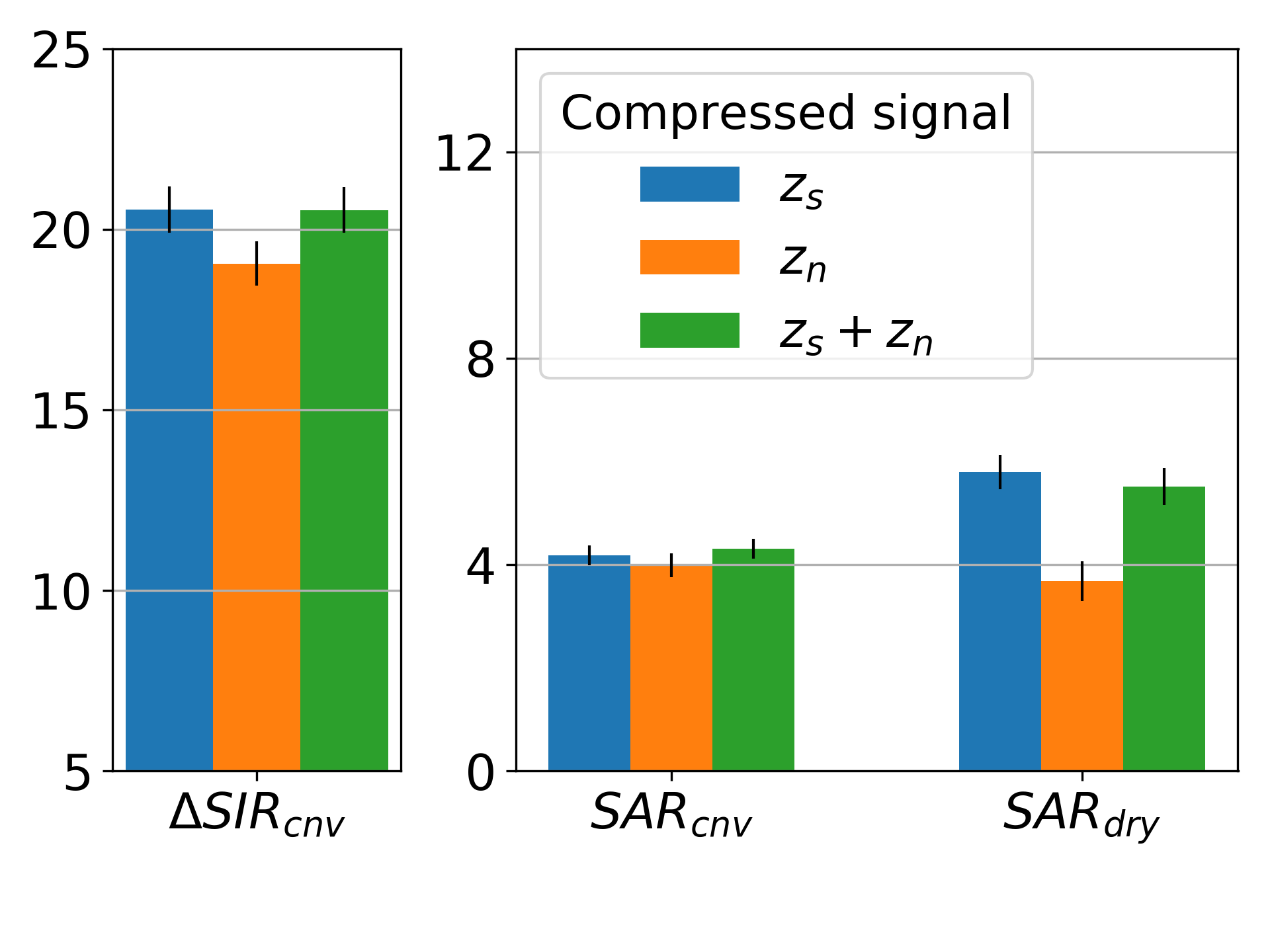}%
		\label{subfig:mn_wi_random}}
	\hfil
	\caption{Speech enhancement performance on the three spatial configurations of multi-node \acp{dnn} trained with different compressed signals.}
	\label{fig:mn_zsn}
\end{figure*}

In this section, we focus on the compressed signal that is sent from one node to the others. In previous versions of \ac{danse}, only the target estimation was sent \cite{Bertrand2010, Hassani2015, Szurley2017, Furnon2020}. However, as depicted in Figure~\ref{fig:sigex}, the noise estimation can also provide useful information, so we propose to compare in which conditions which signal estimation should be sent. To do so, we train three multi-node \acp{dnn}. The first one is the \ac{dnn} that had been used for the previous experiments, and which had as input the target estimations, denoted $z_s$, coming from the distant nodes, together with the noisy signal at the reference channel. The second \ac{dnn} has the noise estimations $z_n$ sent from the distant nodes together with the noisy signal at the reference channel. The third \ac{dnn} has both target speech and noise estimations together with the noisy signal at the reference channel as input. Each of these networks is tested on conditions matching its training conditions, and the results are represented in Figure~\ref{fig:mn_zsn}. We represent the results obtained at the best output nodes (i.e. the result is the average over all the filtered signals obtained at the nodes with the highest output $SIR_{\text{cnv}}$), at the best input node (average over the filtered signals obtained at the nodes with the highest input $SIR_{\text{cnv}}$, and at the worst input node (average over the filtered signals obtained at the nodes with the lowest input $SIR_{\text{cnv}}$)).

At the best output node (Figure~\ref{subfig:mn_bo_random}), sending the one or the other compressed signal does not make any difference. At the best input node (Figure~\ref{subfig:mn_bi_random}), although the differences are not significant, the $\Delta SIR_{\text{cnv}}$ indicates that this node could benefit from receiving the compressed noise estimation rather than the compressed target estimation. Likewise, using the compressed noise estimation at the worst input node (Figure~\ref{subfig:mn_wi_random}) leads to worse results, since the worst input node already has good insights on the noise signal and needs an estimation of the target signal, which it can poorly estimate in its own. 

Sending both the target and the noise estimations seems to be very similar to sending only the target estimation. It looks like it does not benefit from the noise estimation at the best input node. However, the significance of the results allows us only to conclude that sending both estimations is not worse than sending either of both. Given the relatively simple architecture of the network, it could also be that sending both signals from all nodes represents an overload of data for the \ac{dnn}. Carefully selecting either $z_s$ or $z_n$ at the input of the \ac{dnn} might offer a solution to have the best of both worlds, while alleviating the bandwidth requirements.

Hence, depending on the application, if the aim of the speech enhancement challenge is to have the one best signal for the whole microphone array, then sending only $z_s$ is enough. If each node should have its own estimated signal, as discussed in \cite{Markovich-Golan2015}, then depending on the node and its input \ac{sir}, a decision has to be taken whether the target or the noise estimation is of greater relevance. Sending both could is an interesting option but it means sending twice more data.

~\\

Lastly, it is worth noting that the nodes with the best output signal are not always the nodes with the best input signal. The performance of the two filtering steps described in Figure~\ref{fig:tango_nn} at the best input nodes and at the best output nodes is given in Table \ref{tab:s1_vs_s2}. The first (resp. second) filtering step is mentioned as S1 (resp. S2) and the best input node (resp. best output node) is indicated with the subscript $_{\text{bi}}$ (resp. $_{\text{bo}}$). Even at the first filtering step, where the spatial information is not yet shared, the best input nodes are not always the best output nodes, but the difference of performance between the two types of nodes is quite low. The best output nodes benefit more from the second filtering step than the best input nodes. This is because the performance at the second filtering step (where the multi-node \acp{dnn} are used) depends a lot on the compressed signals, which are in general very well estimated by the best input nodes. These compressed signals are received by the other nodes which can benefit from their accuracy and estimate the best output signal. Interestingly, the input $SIR_{\text{cnv}}$ of the best output nodes of the second filtering step (equal to 1~dB) is lower than the input $SIR_{\text{cnv}}$ of the best output nodes of the first filtering step (equal to 1.8~dB). It means that some nodes with a lower input $SIR_{\text{cnv}}$ become the nodes with the best overall performance thanks to the information shared across the microphone array. This phenomenon highlights the cooperation among nodes in the proposed algorithm. 
\begin{table}
	\centering
	\caption{Difference of performance between the first and second filtering steps at the best input node and best output node.}
	\begin{tabular}{|l|c|c|c|c|}
		\hline
		(dB) & $\boldsymbol{SIR_{\text{cnv}}}$ & $\boldsymbol{\Delta SIR_{\text{cnv}}}$ & $\boldsymbol{SAR_{\text{cnv}}}$ & $\boldsymbol{SAR_{\text{dry}}}$ \\
		\hline
		S1$_{\text{bi}}$ & 17.9 $\pm$ 0.4 & 15.3 $\pm$ 0.4 & 7.4 $\pm$  0.2 & 7.6 $\pm$  0.2 \\
		S1$_{\text{bo}}$ & 19.4 $\pm$ 0.4 & 17.6 $\pm$ 0.3 & 7.6 $\pm$  0.1 & 8.0 $\pm$  0.2 \\
		\hline
		S2$_{\text{bi}}$ & 20.5 $\pm$  0.7 & 17.9 $\pm$ 0.5 & 6.4 $\pm$  0.2 & 7.4 $\pm$  0.3 \\
		S2$_{\text{bo}}$ & 23.9 $\pm$  0.5 & 23.0 $\pm$ 0.5 & 6.6 $\pm$  0.2 & 8.3 $\pm$  0.2 \\
		\hline
	\end{tabular}
	\label{tab:s1_vs_s2}
\end{table}

\section{Conclusion}
\label{sec:conclusion}
We introduced and extended a \ac{dnn}-based distributed multichannel speech enhancement methodology which operates in spatially unconstrained microphone arrays. It was evaluated on a large variety of real-life scenarios which proved the efficiency of this solution. It was also shown that this solution is robust to mismatches between the training and test conditions of the \ac{dnn}. We showed that the nodes with the lowest input \ac{sir} benefit the most from the cooperation across the microphone array and we gave insights on the potential benefit of sending the noise estimation rather than the target estimation. To definitely validate the efficiency of this solution, an evaluation on real data remains necessary. 
Another interesting direction of research would be to better select the signals that are needed, either before or after sending them as compressed signals, e.g. with attention mechanisms.

\section*{Acknowledgment}
This work was made with the support of the French National Research Agency, in the framework of the project DiSCogs (ANR-17-CE23-0026-01). Experiments presented in this paper were partially out using the Grid5000 testbed, supported by a scientific interest group hosted by Inria and including CNRS, RENATER and several Universities as well as other organizations (see https://www.grid5000).

\ifCLASSOPTIONcaptionsoff
  \newpage
\fi



\bibliographystyle{../../bib/IEEEtran/IEEEtran}
\bibliography{../../bib/IEEEtran/IEEEabrv,../../bib/disco}

\end{document}